\documentclass[12pt]{iopart}

\usepackage{graphicx}
\usepackage{cite}
\usepackage{caption}
\usepackage{algorithmic}
\usepackage{algorithm}
\usepackage{subfigure}
\usepackage{subcaption}
\usepackage[hyphens]{url}
\usepackage{xltabular}
\usepackage{tabularx}
\usepackage{longtable}
\usepackage{hyperref}

\expandafter\let\csname equation*\endcsname\relax
\expandafter\let\csname endequation*\endcsname\relax
\usepackage{amsmath}

\begin{document}

\title[Quantum Circuit Profiling]{Profiling quantum circuits for their efficient execution on single- and multi-core architectures}

\author{\normalsize{Medina Bandic$^{1,}$$^3$, Pablo le Henaff$^{1,}$$^3$,  Anabel Ovide$^2$, Pau Escofet$^4$, Sahar Ben Rached$^4$,  Santiago Rodrigo$^4$, Hans van\\ Someren$^1$, Sergi Abadal$^4$, Eduard Alarc\'on$^4$, Carmen G. Almudever$^2$, Sebastian Feld$^{1,}$$^3$}}
\address{\footnotesize{Delft University of Technology, The Netherlands$^1$}}
\address{\footnotesize{Universitat Politècnica de València, Spain$^2$}} \address{\footnotesize{QuTech, The Netherlands$^3$}}
\address{\footnotesize{Universitat Polit\`ecnica de Catalunya, Spain$^4$}}


\begin{abstract}
Application-specific quantum computers offer the most efficient means to tackle problems intractable by classical computers. Realizing these architectures necessitates a deep understanding of quantum circuit properties and their relationship to execution outcomes on quantum devices.   
Our study aims to perform for the first time a rigorous examination of quantum circuits by introducing graph theory-based metrics extracted from their qubit interaction graph and gate dependency graph alongside conventional parameters describing the circuit itself. This methodology facilitates a comprehensive analysis and clustering of quantum circuits. Furthermore, it
uncovers a connection between parameters rooted in both qubit interaction and gate dependency graphs, and the performance metrics for quantum circuit mapping, across a range of established quantum device and mapping configurations. Among the various device configurations, we particularly emphasize modular (i.e., multi-core) quantum computing architectures due to their high potential as a viable solution for quantum device scalability. 
This thorough analysis will help us to: i) identify key attributes of quantum circuits that affect the quantum circuit mapping performance metrics;
ii) predict the performance on a specific chip for similar circuit structures; iii) determine preferable combinations of mapping techniques and hardware setups for specific circuits; and iv) define representative benchmark sets by clustering similarly structured circuits.
\end{abstract}

\vspace{2pc}

\maketitle

\section{Introduction}
\label{sec:introduction}
In recent decades, the realm of quantum technology has witnessed remarkable progress, holding the potential to tackle problems that were once deemed insurmountable using classical means. Although these advancements are impressive, we are still in the early stages of understanding its full potential. The current generation of quantum devices, referred to as Noisy Intermediate-Scale Quantum (NISQ) devices \cite{preskill2018quantum}, present severe limitations due to their size and susceptibility to noise. As a result, they are currently adept at handling only simple and modestly-sized circuits (i.e., executable descriptions of algorithms). These devices also face other hurdles, including restricted qubit connectivity, a narrow set of supported operations, and challenges pertaining to classical control resources \cite{Rodrigo2021OnDF,ovide2023mapping}. These collective constraints make the successful execution of a quantum circuit on such processors an intricate endeavor. Furthermore, NISQ devices often adhere to a `one-size-fits-all' approach, which can lead to architectures ill-suited for certain quantum algorithms, resulting in lower success rates and fidelity. This is already well showcased in classical computing, where devices are often tailored for the purpose of usage (e.g., GPUs for gaming).

Most current quantum computers operate as single-processor devices, containing all qubits on a single chip. Scaling these designs proves challenging due to issues like crosstalk and limitations in control electronics \cite{sarovar2020detecting}. An alternative approach, akin to classical computing, involves multi-processor (or multi-core) architectures, which are proposed by various quantum processor manufacturers ~\cite{bravyi2022future,ang2022architectures,Monroe_2014, laracuente2023modeling, https://doi.org/10.48550/arxiv.2201.08861,https://doi.org/10.48550/arxiv.2210.10921}. These new designs facilitate distributed quantum computing that enables the execution of large algorithms across multiple cores to accommodate more qubits than a single processor can handle, and represent a feasible avenue for achieving scalability in quantum computing.

For both NISQ and multi-core architectures, the quantum circuit mapping process \cite{bandic2020structured, baker2020time} is necessary to efficiently run the circuits and maximize the usage of hardware resources. Quantum circuit mapping essentially represents adapting quantum circuits to quantum devices to adhere to all hardware constraints, forming a vital component of a full-stack quantum computing system \cite{bandic2022full}. 

Several studies emphasize the importance of considering a broader range of circuit features during the process of mapping \cite{https://doi.org/10.48550/arxiv.2108.02099,lubinski2021application,mills2020application,li2020towards}. A comprehensive profiling or characterization of quantum circuits offers several advantages, including gaining a better understanding of why certain algorithms achieve higher fidelity with specific processors and mapping techniques \cite{bandic2023interaction}. Additionally, it enables the classification and prediction of the performance of similar circuits based on the circuits' attributes (like in \cite{quetschlich2023predicting}), all without the need for actual hardware execution. Moreover, this approach facilitates the development of mapping techniques and overall quantum systems customized for specific applications, respecting both the requirements (i.e., characteristics) of those particular circuits and the limitations of the hardware \cite{li2021software, lao20212qan, bandic2022full}; this consequently leads to an improvement in quality and execution time of solving currently intractable problems. It is important to note that this exhaustive characterization of quantum circuits is not only the key for formulating meaningful and representative sets of quantum benchmarks that evaluate quantum circuit mapping techniques and entire quantum computing systems \cite{qbench, quetschlich2023mqt}, but also for establishing a suite of algorithm-level metrics to measure system performance \cite{tomesh2022supermarq}.

The state-of-the-art characterization of quantum circuits proposed in \cite{bandic2023interaction} extends beyond conventional metrics such as qubit and gate counts. In addition to these standard attributes, their approach includes an examination of qubit interaction graphs, drawing insights from graph theory and machine learning to clarify the circuit's two-qubit gate connections (qubit interactions). They performed an analysis of how these circuit parameters affect the performance of a specific quantum circuit technique when considering three different single-core devices.  

In this paper, we extend the results of \cite{bandic2023interaction} by additionally encompassing metrics extracted from \textit{gate dependency graphs} (which portray inter-dependencies among gates within the circuit), as well as parameters that describe the \textit{density of the circuit} and its \textit{repetitive oracles}.  
Furthermore, we not only consider parameters that are relevant for single-processor devices but also identify those that are of special interest for the next generation of \textit{multi-core} architectures. 
Within our approach, we also experiment with \textit{diverse quantum circuit mapping configurations} (four for single- and three for multi-core quantum architectures). This intricate exploration allows us to: i) discern the most influential quantum circuit attributes that impact the performance of circuit mapping; ii) predict the mapping performance of similarly structured quantum circuits on a specific chip; iii) identify the most adequate combination of mapping technique and quantum hardware for a given quantum circuit or set of circuits; and iv) define a representative benchmark set \cite{tomesh2022supermarq}, by specifying a finite amount of groups of similarly-structured circuits. 
This thorough analysis, therefore, holds the potential to contribute to the future co-design of compilation methodologies driven by algorithms and the evolution of quantum hardware.

In summary, the main contributions of this paper are:
\begin{enumerate}
    \item Performing the most comprehensive profiling of quantum circuits by extracting: a) standard parameters (i.e., number of qubits and gates, two-qubit gate percentage and depth), metrics from the b) interaction graph (e.g., average node degree), and c) gate-dependency graph (e.g., critical path length), d) gate density related parameters (e.g., amount of idling) and e) characteristics related to repetitive sub-circuits. We believe that this list of parameters encompasses all relevant aspects of a quantum circuit. It helps us gain insights into why certain circuits excel or falter on particular architectures, potentially revealing correlations between circuit structure and performance across different quantum setups. Leveraging these parameters, we can adapt existing or craft new full-stack quantum systems with higher precision.
    \item Identifying, for the first time,  circuit parameters that are key for scalable modular quantum computing architectures. Those architectures demand a unique parameter set due to their intricate quantum circuit mapping requirements, as elaborated in detail in Sec. \ref{Sec2}.
    \item Finding a correlation between extracted circuit features and compilation performance across various mapper-device combinations for single-core and multi-core architectures (totaling in seven mappers and six devices). Utilizing the Pearson correlation score \cite{freedman2007statistics}, we rank parameters from most positively correlated to most negatively correlated for each combination. This analysis shows the significance of selecting suitable device topologies and mapping techniques for quantum circuits with specific structural parameters. It also highlights the key circuit parameters crucial for designing application-aware quantum systems. Identifying these influential circuit parameters marks the initial stride towards crafting such systems.     
    \item Clustering similarly structured circuits and determining the most effective mapper-device setups for them. We illustrate how these clusters also correspond to circuit origin groups found in qbench (i.e., random, QUEKO, real algorithms) \cite{qbench}. This discovery of distinct groups of quantum circuits allows us to establish representative sets of quantum benchmarks without the need for an exhaustive list, which also facilitates the design of application-specific quantum systems tailored to each group, rather than a separate system for each benchmark.
    \end{enumerate}

The paper is organized as follows: Sec. \ref{Sec2} introduces single- and multi-core computation and quantum circuit mapping, as well as the previous work on quantum circuit characterization. Sec. \ref{Sec3} showcases our novel circuit parameters and profiling process in this work, done for single- and multi-core quantum computation separately. In Sec. \ref{Sec4}, we dive into methodology and performance metrics. We show and discuss results in Sec. \ref{Sec5} and finally conclude our work in Sec. \ref{sec:conclusion}.

\section{Background and previous research}
\label{Sec2}

\subsection{Multi-core quantum computation} 

Just like for classical computing, modular architectures are envisioned as a solution for the scalability of quantum devices and they are expected to accommodate thousands to millions of qubits within a single system. For dealing with problems regarding crosstalk, classical control electronics, and wiring complexity, qubits are distributed over multiple cores or processors. This strategy capitalizes on quantum parallelization while addressing qubit control demands and improving qubit isolation \cite{9268630}. Nevertheless, constructing multi-core processors introduces new challenges, mainly due to latency-prone quantum communications that introduce inefficiencies. Quantum state transfer across chip-scale networks represents an alternative to conventional communication technologies, which prove impractical for modular architectures. Communication latencies heighten the risk of data loss during qubit transmission due to state decoherence \cite{Rodrigo2021OnDF}. Several strategies have emerged to create inter-core communication networks for modular quantum computing architectures, accommodating varying technology platforms. These include quantum links for superconducting chips \cite{Bravyi2022thefuture}, ion-shuttling for ion-trapped quantum computers \cite{10.1116/1.5126186}, and photonic networks \cite{Marinelli:2023ojl}. Modularity requirements extend beyond the qubit layer, imposing constraints on networking, control, and compilation layers, necessitating a comprehensive software-hardware stack \cite{9268630}. 

\subsection{The quantum circuit mapping problem}
As previously stated, running a quantum circuit on any device is not a straightforward task. Given the constraints of current NISQ devices (e.g., highly error-prone and limited size and qubit connectivity), the execution of quantum circuits often requires making some modifications within the circuit. This process, referred to as quantum circuit mapping, is pivotal in adapting algorithms to quantum devices, and it forms a vital component of a full-stack quantum computing system \cite{almudever2020realizing}. Limited connectivity is the main challenge addressed by quantum circuit mapping because, during the execution of circuits, all interacting logical/virtual qubits (of the circuit) must be adjacent. This problem is solved by finding the most suitable allocation of logical qubits to physical qubits on the chip, and repositioning those logical qubits on the chip to make them adjacent when necessary, which is usually done by inserting additional gates (e.g., SWAPs or shuttle operations). 
The specifics of quantum circuit mapping can differ from one device to another due to technology disparities and qubit connectivity. Ensuring effective utilization of limited hardware resources and minimizing errors during quantum operation execution by reducing additional gates and circuit latency necessitates this process. Yet, solving the quantum circuit mapping problem as the qubit count increases is computationally challenging, even for current monolithic (or single-core) devices. To address this, diverse quantum circuit mapping algorithms have been introduced, ranging from heuristic and brute-force strategies to graph-theoretical techniques, dynamic programming algorithms, and machine learning-based approaches \cite{zulehner2018efficient, li2019tackling, lao2021timing, itoko2020optimization,pozzi2020using,jiang2021quantum, steinberg1, wagner2023improving,  murali2019noise, tannu2019not, li2020towards, venturelli2019quantum, lao2018mapping, lao2019mapping, herbert2018using,lye2015determining, li2020qubit, biuki2022exact, molavi2022qubit, moro2021quantum, devulapalli2022quantum, upadhyay2022shuttle,nottingham2023decomposing,paraskevopoulos2023spinq, steinberg2024resource}, relying on different performance metrics like gate count, circuit depth (i.e., number of layers of gates where gates can run in parallel), fidelity \cite{murali2019noise,tan2021optimal,tannu2019not} or the circuit success rate \cite{jiang2021quantum,blume2020volumetric}.

\begin{figure*}[h]
\centering
    \includegraphics[width=\linewidth]{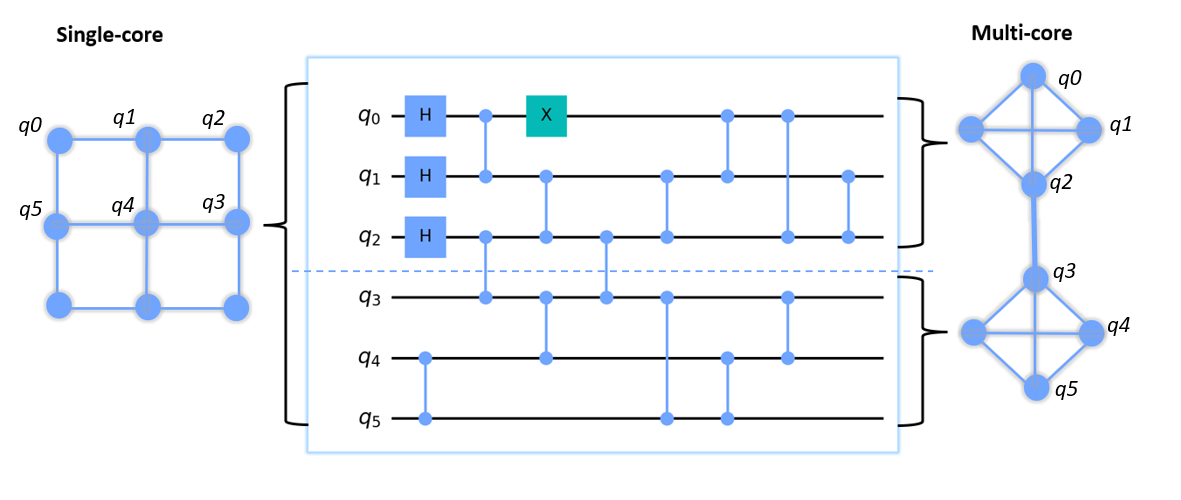}
    \caption{Initial mapping of a quantum circuit (middle) to a single-core (left) or multi-core device (right); For the latter case, the quantum circuit is partitioned into two highly connected circuit slices so that any communication between the two cores is minimized.}
    \label{fig:mapping}
\end{figure*}

However, mapping techniques designed for single-processor NISQ devices do not readily apply to modular multi-core (or multi-node) quantum computing architectures, which offer a promising path for quantum computing scalability \cite{Rodrigo2021OnDF}. This architectural approach involves cores (quantum processing units, or QPUs) interconnected via classical and, ultimately, quantum communication channels. Quantum links facilitate the transfer of quantum states among processors or the execution of inter-core quantum gates depending on the technology, while classical links ensure the coordination of quantum communication \cite{Rodrigo2021}. The intricate communication channels and traffic patterns in multi-core architectures add complexity to quantum circuit mapping compared to single-core devices \cite{rodrigo2022characterizing, ovide2023mapping, rached2023characterizing}. In response, novel techniques have emerged for modular architectures, aiming to minimize the costly (in terms of time and effort overhead or reduced fidelity) long-distance inter-core operations. To reduce the amount of inter-core operations, it is essential to efficiently allocate logical qubits across the physical qubits of the given cores. While the literature in this emerging domain is limited, some approaches have concentrated on quantum compilation and mapping for modular quantum computing \cite{cuomo2023optimized,ferrari2020compiler, bandic2023mappingqubo, escofet2023hungarian}. In these approaches, the quantum circuit is divided into smaller partitions, and interaction graphs reflecting operations within a circuit segment are mapped onto cores by grouping qubits with high interaction levels (see Fig. \ref{fig:mapping}).


All previously stated strategies within both single- and multi-core quantum computation share a common goal: tailoring quantum circuits to device-specific attributes and constraints, while minimizing the communication overhead. However, they often only focus on a limited set of circuit features, such as gate and qubit counts, and qubit interactions. What is missing is a more comprehensive quantum circuit characterization that goes into deeper aspects. For example, one can explore the characteristics of the qubit interaction graph, such as the frequency of interactions between qubit pairs \cite{bandic2023interaction} and the distribution of these interactions among qubits, as well as the quantum instruction dependency graph (representing gate dependencies for scheduling).

Some researchers have already highlighted the significance of incorporating application-specific properties \cite{lubinski2021application, mills2020application, li2020towards, bandic2022full, steinberg1, bandic2023interaction} to enhance quantum circuit mapping and overall quantum system performance. Even in classical computing, the allocation of computing resources depends on the intended applications and processes. In a similar vein, thorough profiling aids in identifying the essential circuit features for successful execution on a particular device and vice versa.
Yet, as we explore the intricacies of running algorithms on modular architectures, it becomes evident that this realm demands a distinct approach when compared to conventional monolithic NISQ devices. Consequently, profiling quantum circuits in the context of modular architectures should adopt a tailored strategy and select parameters that align with the nuances of this scenario.

In conclusion, understanding the structural attributes of quantum circuits sheds light on why and which groups of algorithms perform better on specific processors with designated mapping techniques than on others. This holistic understanding not only guides mapping but also opens avenues for improving the overall performance of quantum circuits on quantum hardware and represents a first step towards application-based quantum computers.

\subsection{On the importance of qubit interaction and gate-dependency graphs}

Quantum circuits so far have mostly been characterized in terms of size, i.e., the number of qubits, gates, two-qubit gates, and depth, which are blind to the circuit's structure. Considering that the main quantum hardware constraints are low fidelity and limited qubit connectivity, it is important to extract more information about the qubit interaction distribution as well as gate dependencies as they directly relate to gate and depth overhead that results from the compilation process. Previous works have emphasized the significance of deriving additional circuit parameters based on qubit interaction and quantum instruction dependency graphs for the development of mapping techniques \cite{li2019tackling,lao2021timing,baker2020time,bandic2023mappingqubo}. While gate dependency graphs have been utilized for operation scheduling optimization and look-ahead techniques, interaction graphs have typically been employed for the initial qubit placement and routing procedure. In this work, we utilize these two graphs and their graph-theory-based attributes for characterizing circuits targeting both single- and multi-core architectures. In the next paragraphs, we will introduce these two graph representations of the quantum circuit.

\begin{figure}[h!]
    \centering
    \centerline{
\subfigure[]{\includegraphics[width=0.3\linewidth]{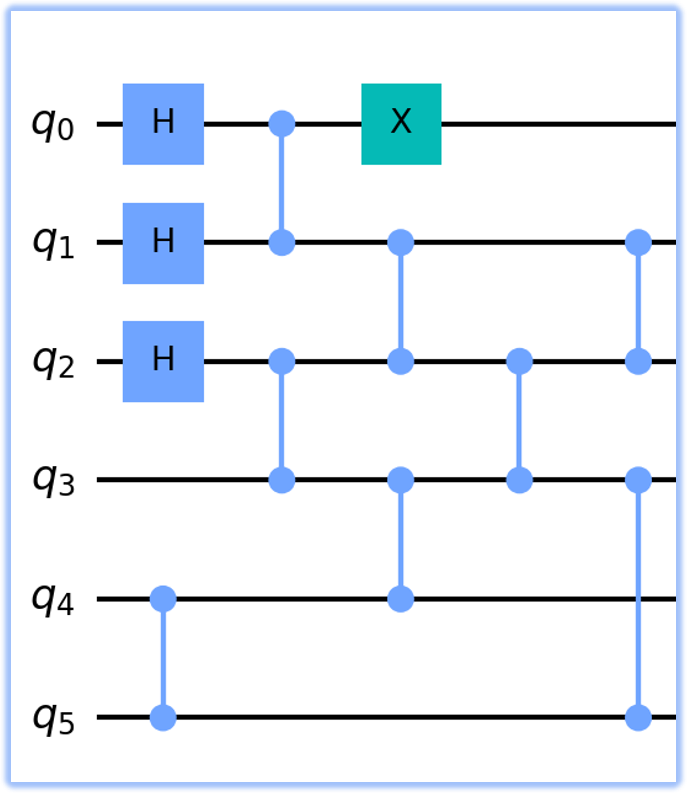}
\label{fig:qc}}
\subfigure[]{\includegraphics[width=0.35\linewidth]{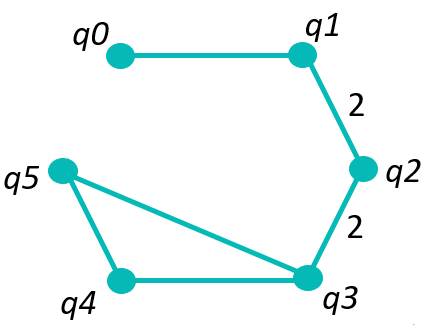}
\label{fig:ig}}
\subfigure[]{\includegraphics[width=0.35\linewidth]{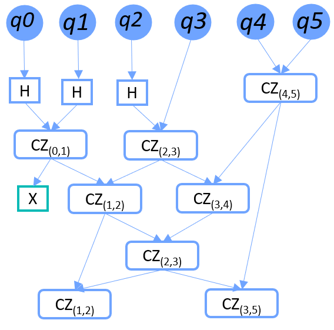}
\label{fig:gdg}}}
\caption{An exemplary quantum circuit (a) and its corresponding qubit interaction graph (b) and gate dependency graph (c).}
\label{fig:ig_gdg}
\end{figure}

The \textit{ Interaction Graph} (IG) $G_i(V_i, E_i)$ offers a visual representation of the spatial distribution of two-qubit gates for a given quantum circuit. Typically, this is an undirected and connected graph (see Fig. \ref{fig:ig}), with edges denoted as $e(v_i,v_j) \in E_i$ representing the two-qubit gates and nodes labeled as $v_i \in V_i$ representing the qubits engaged in these gates \cite{bandic2023interaction,bandic2023mappingqubo}. Previous research \cite{bandic2023interaction} has demonstrated that different categories of circuits (e.g., real algorithms and random circuits) exhibit varying compilation performance in terms of gate and depth overhead as well as fidelity decrease when executed on different physical-qubit topologies (see Figs. 5 and 14 of \cite{bandic2023interaction}). To understand the underlying reasons for these differences, prior studies have explored the structural disparities between various quantum circuit groups based on their interaction graphs. The quantum circuit mapping problem can be conceptualized as a graph problem, in which the comparison between the interaction graphs of a circuit and the coupling graph of the device is crucial \cite{steinberg2024resource}. Therefore, the parameters of the interaction graph can significantly influence the outcome. Studies such as \cite{tomesh2022supermarq, bandic2023interaction} have illustrated how the interaction graph and size parameters of quantum circuits directly correlate to their performance on different chips when employing the same mapping technique. For example, circuits with fully connected interaction graphs are expected to incur higher overhead, unless the processor's coupling graph is fully or almost fully connected. Moreover, research such as \cite{bandic2023interaction} has further clustered the circuits based on these extracted parameters and identified performance discrepancies among the groups based on the aforementioned mapping metrics.

Interaction graphs play an even more significant role in the compilation of circuits for multi-core systems \cite{escofet2024revisiting}. Mapping to multi-core devices involves minimizing inter-core communication by partitioning the interaction graphs and effectively allocating the logical qubits (IG nodes) onto the physical qubits of different cores. This is achieved by aiming to map highly interacting qubits (with high IG weights) onto the same core. Given the importance of IG partitioning in modular compilation and its distinct approach compared to single-core systems, it is essential to use different IG metrics. These metrics should focus on identifying local clusters and cliques to effectively describe how 'partitionable' a circuit is.

Besides that, the \textit{Gate-Dependency Graph} (GDG) 
$G_{gd}(V_{GD}, E_{GD})$ is a directed and connected graph illustrating the interrelations between the gates in the circuit. Each gate in the circuit is represented by a node $v_j \in V_{GD}$, and the dependency between gates is depicted as a directed edge $e(v_j,v_k) \in E_{GD}$. An illustrative example of this concept is shown in Fig. \ref{fig:ig_gdg}, where a quantum circuit consisting of 6 qubits and 8 two-qubit gates is displayed alongside its corresponding IG and GDG \cite{lao2019mapping}. It is worth noting that both graphs can also be represented as weighted graphs: the IG becomes weighted when multiple two-qubit operations occur between the same pairs of qubits, while the GDG becomes weighted when considering, for instance, gate duration (i.e. time). For the purpose of this paper, we focus on the simplified, unweighted version of GDG, where all gates take one time step.


In this paper, we aim to examine quantum circuits by analyzing parameters extracted from interaction graphs that are relevant for single- and multi-processor architectures, from gate dependency graphs (e.g., path distribution and critical path length), parameters related to repeating sub-circuits/oracles within the circuit, and circuit density-related parameters, followed by a clustering of the circuits. Our focus is to identify and quantify the relationships between circuit parameters, clusters, and various compilation techniques across different quantum topologies. Gaining a comprehensive understanding of circuit structures can aid in designing quantum systems optimized for the highest success rates for specific circuit groups and help define a finite set of representative benchmarks.  

In the subsequent sections, we will provide a detailed overview of the entire quantum circuit profiling process.

\section{Characterizing quantum circuits -- definition and profiling}
\label{Sec3}

\subsection{Parameter selection}

Tab. \ref{tab:circ_params} illustrates the five groups of circuit parameters we consider in this work. The process selection of the \textbf{parameters of interest } involved two steps: 1) Selecting parameters within these groups relevant to single- and multi-core architectures considering their main constraints. For instance, the clustering coefficient \cite{hernandez2011classification} is important for the execution of circuits in multi-core devices but not in single-core devices, as it requires graph partitioning. Conversely, node degree is relevant to both scenarios: a higher average degree in the interaction graph makes mapping more challenging; and 2) Reducing the number of parameters by identifying highly correlated ones using the Pearson correlation matrix \cite{freedman2007statistics}, like in \cite{bandic2023interaction}.

The resulting reduced set of parameters is as follows \cite{hernandez2011classification,bandic2023interaction}:

\subsubsection{\textbf{Size}}
   Standard parameters used for circuits description in previous works: \begin{itemize}
        \item \textit{Number of qubits}: $n_q$
        \item \textit{Number of gates}: $n_g$
        \item \textit{Two-qubit gate percentage}: $\frac{n_{2qg}}{n_g},$ 
        where $n_{2qg}$ is the number of two-qubit gates.
        \item \textit{Decomposed circuit depth}: $d$
    \end{itemize}
\subsubsection{\textbf{IG} (definitions taken from \cite{bandic2023interaction})}
\label{Sec3.1.2}

    \begin{itemize} 
        \item \textit{Average shortest path length}: Average hopcount between all nodes \cite{hernandez2011classification}. 
        The larger the average hopcount
between the nodes, the less
connected the graph is. It also means that a simpler
interaction graph is easier to
map. 
        \item \textit{Standard deviation of adjacency matrix} $\sigma (A)$: An adjacency matrix $A$ is a square matrix used for representing a graph whose elements are $a_{ij} = a_{ji}$ represent the number of connections between nodes $n_i$ and $n_j$. It shows which nodes are connected and with how many edges. A large $\sigma$ value means some specific pairs of qubits interact much more than others and that there is less additional routing required.  
        \item \textit{Diameter}: Longest shortest path in the circuit, 
        $$dm = \max_{n_i \in \mathcal{N}}(\epsilon_i),$$
        where $\epsilon_i$ is the longest hopcount between node $n_i$ and any other node among the total of $N$ nodes.
        The larger the diameter, the simpler the IG and, therefore, easier to map onto a device.
        \item \textit{Central point of dominance}: Maximal betweenness of any node in the graph, where betweenness is the number of shortest paths between nodes that
traverse some node or edge \cite{hernandez2011classification}. A value of 0 results for complete graphs, and 1 for star-shaped graphs. 
Values approaching $0$ or $1$ are undesirable from the perspective of quantum circuit mapping, as $0$ reports a graph that is too much connected, and $1$ indicates that one qubit is involved in all gates, making the circuit hard to parallelize. 
\item \textit{Average degree}: Average degree of neighbor nodes, where the degree is the number of nodes to which one node is connected and defined as $$deg_i = \sum_{j=1}^N a_{ij}.$$
         The lower the value of the average degree, the less connected the IG is, and the easier it is to map.  
        \item \textit{Number of maximal cliques}: The total number of the largest all-to-all connected subgraphs. This metric also depends on the size of the maximal clique. The smaller the largest clique, the less connected the graph and, therefore, easier to map. 
        
        Another cliques-related graph metric is \textit{clustering coefficient} which measures the cliquishness of a neighborhood. The values range between $0$ and $1$, where $1$ represents a fully connected graph, which is always the worst-case scenario for the quantum circuit mapping:
        $$c_i = \frac{y_i}{\binom{deg_i}{2}},$$ 
        where $y_i$ is the number of links between neighbors of node $n_i$ 
         These two metrics are of high importance for multi-core computation as they show the presence of highly connected clusters of nodes within the graph, which is related to the IG partitioning part of multi-core quantum circuit mapping.
         
         \item \textit{Vertex/edge reliability}: The minimal number of nodes/edges whose removal can disconnect the graph. The lower the reliability, the easier it is to partition the graph for multi-core mapping.
         \item \textit{Coreness}: Maximal $k$ for specific node $i$ such that $i$ is present in $k$-core graph but removed from $(k+1)$-core ($k$-core is a subgraph of some graph made by removing all the nodes of degree $<=k$). Coreness as the local metrics also relates to IG partitioning and modular computing.
         \item \textit{Pagerank}: Ranking of the importance of each node in the graph \cite{pagerank} based on the number and weights of the links with other nodes and the rank of those nodes. This graph metric emphasizes which nodes should be mapped to the most connected part of the chip.    
    \end{itemize}
    
\subsubsection{\textbf{GDG}}

One of the main disadvantages of current QPUs is the short lifetime of qubits, i.e., their decoherence. That makes the circuit \textit{scheduling} one of the crucial segments of the quantum circuit compilation. The goal during scheduling is to make the gates run as parallel as possible. However, not every circuit is parallelizable. GDG and especially its \textit{critical path} (longest path in GDG) showcases the minimal necessary duration of the circuit and represents its longest inter-depending gate sequence. Therefore, this sequence of gates should be scheduled to run as soon as possible in order to shorten the circuit duration and prevent qubit decoherence. Metrics related to critical and other paths of GDG can also tell us how sequential the circuit is and, therefore, how easy it is to map to a single or to multiple cores. Qubits participating in the inter-dependent gates (gates of the same common paths) should be placed nearby on a chip or within the same core during the mapping process. Furthermore, the higher the percentage of the circuit gates included in the critical path, the more sequential the circuit is and, therefore, less parallelizable. Metrics that describe the GDG paths include:
\begin{itemize}
    \item \textit{Critical path length} or number of gates in the critical path;
    \item \textit{Number of critical paths};
    \item \textit{Path length distribution} and its mean and standard deviation; and
    \item \textit{Percentage of gates included in the critical path}.
\end{itemize} These metrics' implementation and detailed definitions are shown in \ref{app:2}. 

\subsubsection{\textbf{Circuit density}}
This set of metrics evaluates the degree of parallelization of the circuit gates before any optimizations. It indicates the number of gates executed in each layer (time-step) of the circuit relative to the maximum number of gates that could be executed if the circuit were fully parallelized (similar to Quantum Volume circuits \cite{blume2020volumetric}). A denser circuit implies greater difficulty in further optimization and execution. This paper utilizes two metrics to describe this behavior:
\begin{itemize}
    \item Density score: Parallelization level of the circuit;
    $$ D = \frac{\frac{2*n_{2qg} + n_{1qg}}{d}-1}{n_q-1},$$ where $n_{2qg}$ and $n_{1qg}$ are number of two- and single-qubit gates, respectively. This is an extended version of the parallelism metric from \cite{tomesh2022supermarq}: we made a distinction between the single- and two-qubit gates for better preciseness, where $D$ can actually reach all the values in the range between $0$ and $1$, ($1$ is maximal density); and 
    \item Idling score: Average amount of qubit idling in the circuit; $$I = \frac{\sum_{i=1}^{n_q} d - q_i}{n_q*d},$$ where $q_i$ signifies the number of layers of the circuit in which the qubit is used; range between $0$ and $1$, with $0$ meaning no idling and $1$ meaning no scheduled gates.
\end{itemize}
\subsubsection{\textbf{Longest sub-circuit repetitions}}
In circuits that are based on real algorithms, there are always patterns in terms of gate order and repetitions. In contrast, completely random circuits show, on average, no such patterns. In order to express the randomness of the gates and groups of gates in circuits, we have defined the following metrics
\begin{itemize}
    \item The number of occurrences of the largest repetitive sub-circuit; and
    \item The size of this largest repetitive sub-circuit.
\end{itemize}
The extraction of these two metrics relies on existing algorithms used for strings of characters. The problem of finding the longest repeating sub-string in a text and the length thereof is efficiently solved by filling a data structure called a \textit{suffix tree}\cite{suffixtreeswiki}. We used the same implementation by identifying characters and quantum gates.
The significance of gate randomness on circuit performance is closely tied to the type of mapper employed. Some mappers are expected to exhibit a high correlation, where the routing algorithm is designed to identify gate patterns in advance. In contrast, more stochastic approaches do not benefit from recognizing these patterns.

The IG parameters reveal patterns necessary for initial placement algorithms, whereas GDG parameters indicate the length and sequential nature of the circuit, crucial for mitigating decoherence and enhancing scheduling and routing. Parameters related to circuit density illustrate the extent of parallelization in the circuit before optimization, influencing the complexity of all stages of quantum circuit mapping. Additionally, identifying recurring gate patterns in the circuits assists in refining scheduling and look-ahead routing techniques. For the complete list of metrics, please refer to \cite{bandic2020structured} (IG-based) and \ref{app:4} and \ref{app:3}.

\begin{table*}[h!]
\caption{Selected metrics for the characterization of quantum circuits.}
\begin{tabular}{|l|l|l|}
\hline
\multicolumn{1}{|c|}{\textbf{Metric}}& \multicolumn{1}{c|}{\textbf{Metric Type}}&   \multicolumn{1}{c|}{\textbf{Single- or Multi-core}} \\         \hline
\begin{tabular}[c]{@{}l@{}}Num. of qubits\end{tabular}                   & \begin{tabular}[c]{@{}l@{}}Size\end{tabular}                      & \begin{tabular}[c]{@{}l@{}}Both \end{tabular} \\
\hline
\begin{tabular}[c]{@{}l@{}}Num. of gates\end{tabular}                   & \begin{tabular}[c]{@{}l@{}}Size\end{tabular}                      & \begin{tabular}[c]{@{}l@{}}Both \end{tabular} \\
\hline
\begin{tabular}[c]{@{}l@{}}Two-qubit gate \%\end{tabular}                   & \begin{tabular}[c]{@{}l@{}}Size\end{tabular}                      & \begin{tabular}[c]{@{}l@{}}Both \end{tabular} \\
\hline
\begin{tabular}[c]{@{}l@{}}Circuit depth\end{tabular}                   & \begin{tabular}[c]{@{}l@{}}Size\end{tabular}                      & \begin{tabular}[c]{@{}l@{}}Both \end{tabular} \\
\hline
\begin{tabular}[c]{@{}l@{}}Avg. shortest path\end{tabular}                   & \begin{tabular}[c]{@{}l@{}}IG\end{tabular}                      & \begin{tabular}[c]{@{}l@{}}Single-core \end{tabular} \\

\hline
\begin{tabular}[c]{@{}l@{}}Standard deviation of adjacency matrix \end{tabular}                   & \begin{tabular}[c]{@{}l@{}}IG\end{tabular}                      & \begin{tabular}[c]{@{}l@{}}Single-core \end{tabular} \\

\hline
\begin{tabular}[c]{@{}l@{}}Diameter\end{tabular}                   & \begin{tabular}[c]{@{}l@{}}IG\end{tabular}                      & \begin{tabular}[c]{@{}l@{}}Multi-core \end{tabular} \\

\hline
\begin{tabular}[c]{@{}l@{}}Central point of dominance\end{tabular}                   & \begin{tabular}[c]{@{}l@{}}IG\end{tabular}                      & \begin{tabular}[c]{@{}l@{}}Multi-core \end{tabular} \\

\hline
\begin{tabular}[c]{@{}l@{}}Maximal cliques and num. of maximal cliques\end{tabular}                   & \begin{tabular}[c]{@{}l@{}}IG\end{tabular}                      & \begin{tabular}[c]{@{}l@{}}Multi-core \end{tabular} \\
\hline
\begin{tabular}[c]{@{}l@{}}Clustering coefficient\end{tabular}                   & \begin{tabular}[c]{@{}l@{}}IG\end{tabular}                      & \begin{tabular}[c]{@{}l@{}}Multi-core \end{tabular} \\
\hline
\begin{tabular}[c]{@{}l@{}}Avg. degree\end{tabular}                   & \begin{tabular}[c]{@{}l@{}}IG\end{tabular}                      & \begin{tabular}[c]{@{}l@{}}Both \end{tabular} \\

\hline
\begin{tabular}[c]{@{}l@{}}Vertex/edge reliability\end{tabular}                   & \begin{tabular}[c]{@{}l@{}}IG\end{tabular}                      & \begin{tabular}[c]{@{}l@{}}Multi-core \end{tabular} \\

\hline
\begin{tabular}[c]{@{}l@{}}Coreness\end{tabular}                   & \begin{tabular}[c]{@{}l@{}}IG\end{tabular}                      & \begin{tabular}[c]{@{}l@{}}Multi-core \end{tabular} \\
\hline
\begin{tabular}[c]{@{}l@{}}Pagerank\end{tabular}                   & \begin{tabular}[c]{@{}l@{}}IG\end{tabular}                      & \begin{tabular}[c]{@{}l@{}}Multi-core \end{tabular} \\
\hline
\begin{tabular}[c]{@{}l@{}}Critical path length\end{tabular}                   & \begin{tabular}[c]{@{}l@{}}GDG\end{tabular}                      & \begin{tabular}[c]{@{}l@{}}Both \end{tabular} \\
\hline
\begin{tabular}[c]{@{}l@{}}Num. of critical paths\end{tabular}                   & \begin{tabular}[c]{@{}l@{}}GDG\end{tabular}                      & \begin{tabular}[c]{@{}l@{}}Both \end{tabular} \\
\hline
\begin{tabular}[c]{@{}l@{}}GDG path length distribution metrics\end{tabular}                   & \begin{tabular}[c]{@{}l@{}}GDG\end{tabular}                      & \begin{tabular}[c]{@{}l@{}}Both\end{tabular} \\
\hline
\begin{tabular}[c]{@{}l@{}}\% of gates in critical path\end{tabular}                   & \begin{tabular}[c]{@{}l@{}}GDG\end{tabular}                      & \begin{tabular}[c]{@{}l@{}}Both \end{tabular} \\
\hline
\begin{tabular}[c]{@{}l@{}}Density score\end{tabular}                   & \begin{tabular}[c]{@{}l@{}}Circuit density\end{tabular}                      & \begin{tabular}[c]{@{}l@{}}Both \end{tabular} \\
\hline
\begin{tabular}[c]{@{}l@{}}Idling score\end{tabular}                   & \begin{tabular}[c]{@{}l@{}}Circuit density\end{tabular}                      & \begin{tabular}[c]{@{}l@{}}Both \end{tabular} \\
\hline
\begin{tabular}[c]{@{}l@{}}Num. of largest rep. sub-circuit\end{tabular}                   & \begin{tabular}[c]{@{}l@{}} Sub-circuit repetitions\end{tabular}                      & \begin{tabular}[c]{@{}l@{}}Both \end{tabular} \\
\hline
\begin{tabular}[c]{@{}l@{}}Size of largest rep. sub-circuit\end{tabular}                   & \begin{tabular}[c]{@{}l@{}} Sub-circuit repetitions\end{tabular}                      & \begin{tabular}[c]{@{}l@{}}Both \end{tabular} \\
\hline
\end{tabular}
\label{tab:circ_params}
\end{table*}

\subsection{Quantum circuit clustering}
\label{sec:3.2}

As previously stated, one of our objectives is to identify structural similarities among quantum circuits and establish ``circuit families'', wherein the constituent elements (i.e., the quantum circuits) exhibit similar compilation behavior and require comparable hardware resources. 
For this purpose, we employed a two-step clustering approach: an initial clustering of circuits based on size parameters (Group 1), followed by clustering based on the remaining parameters. This strategy was implemented to prevent size parameters from exerting undue influence on the clustering algorithm. Consequently, we initially categorized the set of $341$ selected benchmarks (see Sec. \ref{Sec4}) into five clusters using the K-means clustering algorithm \cite{1056489}. Subsequently, each of the five size-related clusters could be further subdivided into sub-clusters based on the previously described structural circuit parameters (groups 2-5). In this regard, we once again opted for the K-means algorithm after evaluating various methods and parameter configurations using the silhouette coefficient method \cite{rousseeuw1987silhouettes}. The specific settings for clustering single- and multi-core devices, as well as the clusters themselves, are detailed or referred to in \ref{app:3} and \ref{app:4}, respectively. It is expected that circuits assigned to the same sub-cluster will exhibit similar fidelity and gate overhead outcomes. The correlation between our circuit groups and mapping performance metrics, as well as potential explanations for these relationships, will be discussed in the subsequent sections.

\section{Methodology}
\label{Sec4}

In our research, the quantum circuit profiling process encompasses the following four steps:

\begin{enumerate}
    \item \textbf{Benchmark collection} -- We gather benchmarks (i.e., quantum circuits) from diverse sources, translate them into the same quantum language (in our case cQASM \cite{qinspire}), and extract their interaction and gate-dependency graphs \cite{bandic2023interaction}. 
    \item \textbf{Parameter selection and extraction} -- We select and extract graph-theory-based parameters from the qubit interaction graph and gate dependency graph, focusing on parameters relevant to mapping quantum circuits to single-core and multi-core devices. Additionally, we extract supplementary circuit parameters to enhance the characterization related to circuit density. 
    \item \textbf{Benchmark clustering} -- We cluster benchmarks based on their size-related (number of qubits and gates, two-qubit gate percentage, and circuit depth) and structure-related parameters.
    \item \textbf{Compilation} -- We compile the quantum circuits using Qiskit \cite{Qiskit}, OpenQl \cite{khammassi2021openql}, and additional multi-core compilation solutions \cite{bandic2023mappingqubo,escofet2023hungarian}, and analyze the relationship between their performance and the extracted parameters, as well as their cluster affiliation (see Sec. \ref{Sec5}).
\end{enumerate}

\subsection{Quantum benchmarks selection}
\label{sec:4.1}
In this paper, we employed the qbench benchmark set \cite{qbench, bandic2023interaction}, which offers a comprehensive collection of quantum circuits sourced from various platforms and written in different programming languages. This set encompasses a range of circuit types categorized based on their origin, including circuits derived from real quantum algorithms (e.g., QFT), simpler algorithm-based circuits (such as arithmetic circuits), QUEKO circuits optimized for specific devices \cite{Queko}, and randomly generated circuits produced by randomly selecting single- and two-qubit gates from a predefined set and applying them to arbitrarily chosen qubits or qubit pairs \cite{DiogoRandomCirc}. This category also includes highly parallelized Quantum Volume circuits used for device benchmarking \cite{cross2019validating}. 

Given the inclusion of multi-core architectures in our study, we also incorporated benchmarks used so far for multi-core computations \cite{rached2023characterizing,ovide2023mapping,bandic2023mappingqubo, baker2020time}. These benchmarks comprise a variety of circuit instances, ranging from 16 to 128 qubits, such as the Cuccaro Adder \cite{article_cuccaro_adder}, Grover's main routine, GHZ state preparation \cite{greenberger2007going}, QFT, QAOA, and VQE \cite{nielsen2002quantum}. To accommodate this aspect, we expanded our synthetic circuit set to include instances with up to 128 qubits. The complete list of 341 quantum circuits utilized in our study is detailed in \ref{app:1}, with accompanying code available at \cite{qbench}.


\subsection{Hardware configuration}
\label{Sec:4.2}
\begin{figure*}[h!]
    \centering
    \centerline{
\subfigure a){\includegraphics[width=0.23\linewidth]{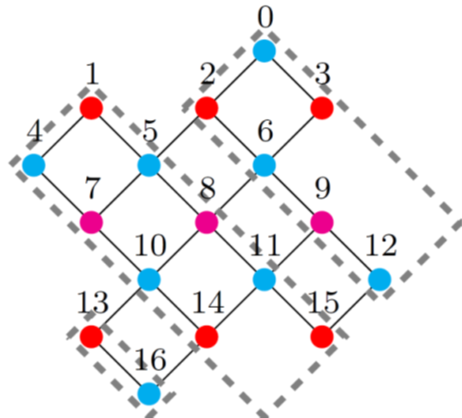}
\label{fig:surf17}}
\subfigure b){\includegraphics[width=0.23\linewidth]{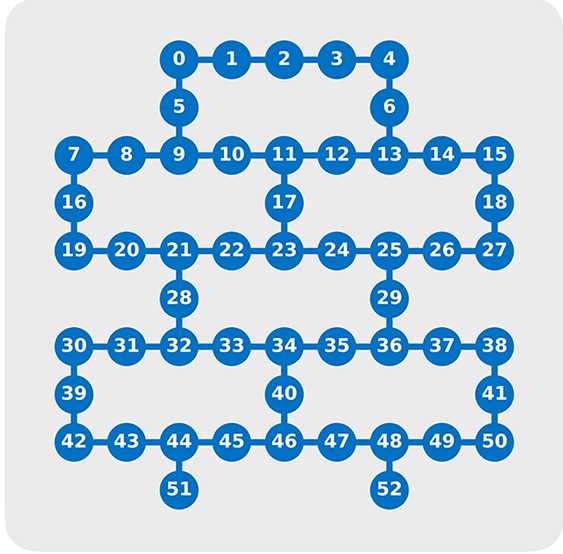}
\label{fig:ibm53}}
\subfigure c){\includegraphics[width=0.23\linewidth]{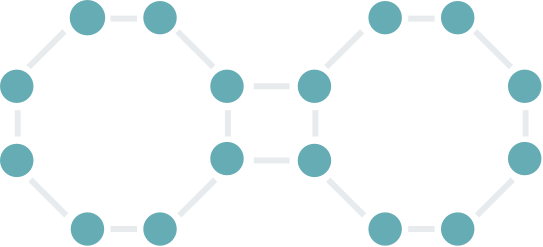}
\label{fig:asp16}}
\subfigure d){\includegraphics[width=0.23\linewidth]{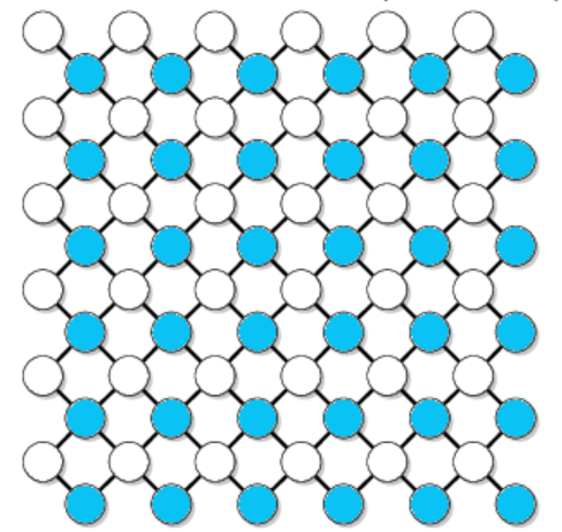}
\label{fig:google72}}}
\caption{Topologies of the quantum architectures used in our experiments: a) Surface-17; b) IBM Rochester; c) Rigetti 16-q Aspen and d) Google Bristlecone.  Figures and device configurations taken from: \cite{lao2021timing, IBM, Rigetti, Goodrich2018PracticalGB}.}
\label{fig:scdevices}
\end{figure*}

\begin{figure*}[h!]
	\centering
	\includegraphics[width=0.5\linewidth]{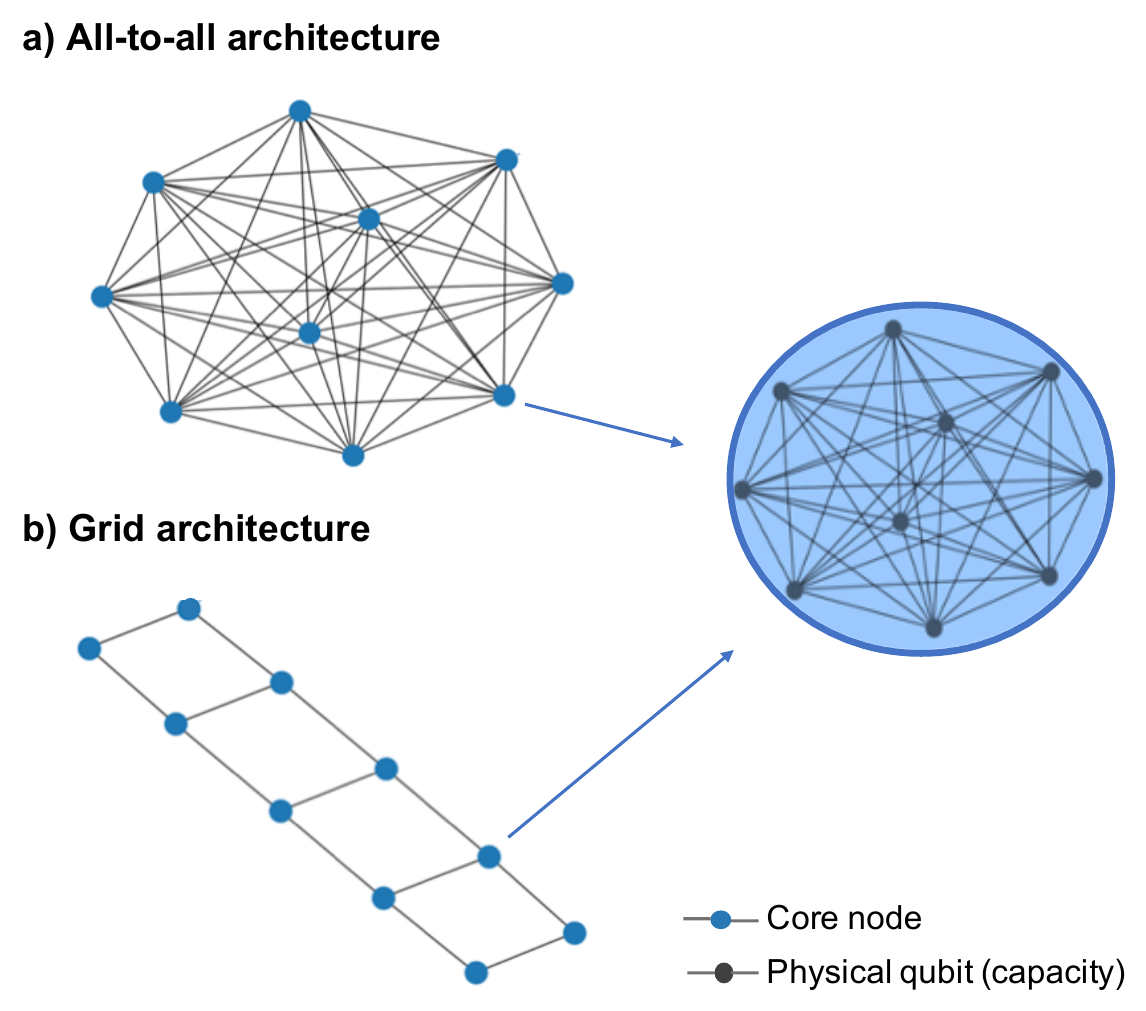}
 	\caption{In our experiments, we examined two distinct multi-core architectures: \textbf{a)} All-to-all connected cores and \textbf{b)} 2D Grid core connectivity. Each node in the two graphs on the left represents a core, while the edges correspond to communication links between the cores. On the right, the intra-core qubit topology is displayed, comprising 10 all-to-all connected qubits. We chose these topologies due to their ability to offer a simplified representation of architectures anticipated to emerge in the near future. This approach enables us to promptly address the immediate need for mapping solutions based on minimizing inter-core communication before going into more complex architectures \cite{bandic2023mappingqubo}.}
 	\label{fig:mcdevices}
\end{figure*}

To investigate the relationship between the previously identified circuit clusters (see Sec. \ref{Sec3}) and the mapping outcomes, we compiled the selected quantum circuits using various target single-core and multi-core quantum architectures. For single-core architectures, we utilized the \textit{Rigetti Aspen-1}, \textit{Surface-17} \cite{lao2021timing}, \textit{IBM Rochester}, and \textit{Google Bristlecone} devices, as depicted in Fig. \ref{fig:scdevices}. Regarding modular architectures, we adopted the same configuration as detailed in \cite{bandic2023mappingqubo}: all-to-all connected qubits within the cores, while the cores themselves are interconnected either in an all-to-all manner or in a grid-like fashion, as illustrated in Fig. \ref{fig:mcdevices}. These device configurations were selected because they are commonly employed in quantum circuit mapping research and offer realistic and diverse connectivity patterns in their coupling graphs. Note, that we did not execute the quantum circuits on actual devices; instead, the hardware constraints of the devices were considered during the compilation process alone.

\subsection{Quantum compilers}
\label{Sec:4.3}
In this study, we investigate not only the relationship between different quantum circuit groups, parameters, and quantum devices but also explore the connection between quantum circuits and mapping techniques. The goal is to gain insights into developing and selecting mapping techniques tailored to circuits with specific common properties.

For single-core devices, we utilize the widely used Qiskit compiler and test four combinations of compilation settings by pairing two routers with two optimization levels. Specifically, we employ the \textit{Stochastic} and \textit{Sabre} routers, along with optimization levels 1 and 2. The key distinction between the Stochastic and Sabre approaches lies in their routing path selection: Sabre utilizes a more deterministic approach, whereas Stochastic employs a more randomized strategy. Optimization level 1 involves simplistic mapping with minimal circuit optimization but short runtime, while optimization level 2 employs more aggressive gate cancellation and commutation techniques to minimize qubit crosstalk and increase parallelism. Level 2 is suitable for complex circuits or cases in which circuit optimization is crucial despite requiring more computational resources and time during compilation. For simplicity, we will refer to these four method combinations as \textit{Stochastic1}, \textit{Stochastic2}, \textit{Sabre1}, and \textit{Sabre2}.

In the modular regime, we explore three techniques:

 \begin{itemize}
     \item The \textit{time-sliced circuit partitioning} method \cite{baker2020time}, which leverages qubit clustering to create tractable partitioning heuristics for mapping quantum circuits to modular physical machines one slice at a time. This method also uses a tunable lookahead scheme to reduce the cost of moving to future time
slices.
     \item The \textit{Hungarian Qubit Assignment} (HQA) algorithm \cite{escofet2023hungarian}, which employs the Hungarian algorithm \cite{kuhn1955hungarian} to enhance qubit-to-core assignment by considering interactions between qubits across the entire circuit,  and enabling fine-grained
partitioning and enhanced qubit utilization. 
     \item A \textit{QUBO}-based approach \cite{punnen2022quadratic}, which encodes qubit allocation and inter-core communication costs in binary decision variables \cite{bandic2023mappingqubo}. This method splits the quantum circuit into slices and formulates the qubit assignment as a graph partitioning problem for each slice, reducing costly inter-core communication by penalizing that. The final solution minimizes the overall cost across all circuit slices.
 \end{itemize}

\subsection{Performance metrics}

\textbf{Quantum circuit mapping performance} metrics are defined as follows:

\begin{enumerate}
    \item \textbf{Gate overhead} is calculated using the formula: $G_{overhead} = \frac{(G_{after} - G_{before)}}{G_{before}}$. Here, $G_{before}$ and $G_{after}$ denote the number of gates before and after compilation, respectively. 
    \item \textbf{Depth overhead} is determined by:
    $D_{overhead} = \frac{(D_{after} - D_{before})}{D_{before}}$, where $D_{before}$ and $D_{after}$ represent the circuit depth before and after compilation. Depth refers to the number of cycles or layers of simultaneously running gates in the circuit. 
    \item \textbf{Fidelity decrease} is computed as:
    $F_{decrease} = \frac{(F_{before} - F_{after})}{F_{before}}$.
    Here, $F_{before}$ and $F_{after}$ represent the circuit fidelity before and after compilation, respectively. \textit{Circuit fidelity} is a product of the error rates of the gates in the circuit. The primary objective during circuit mapping is to maximize this metric. We assumed uniform error rates for all one-qubit and two-qubit gates, using average values from the Starmon-5 chip \cite{murali2019full,nishio2020extracting,qinspire}.
    
    \item \textbf{Number of non-local (inter-core) communications} is the number of qubit 
    moves from one core to another within modular quantum architectures. Entanglement-based quantum communication protocols are employed that require Bell-pair generations that enable the teleportation of quantum states. However, generating entangled pairs results in a resource overhead, and the process itself is non-deterministic, adding complexity to scheduling tasks \cite{ovide2023mapping}.
    
\end{enumerate}



In the following section, we will discuss the relation of the structural parameters of circuits from Sec. \ref{sec:4.1} with the above-stated obtained metrics after mapping them into different combinations of quantum system setups mentioned in Secs. \ref{Sec:4.2} and \ref{Sec:4.3}.  

\section{Results and discussion}
\label{Sec5}

\label{sec:results}

In this section, we assess and contrast the mapping results of our chosen circuits while examining the influence of circuit parameters on the outcomes. Furthermore,
we juxtapose the performance of various circuit clusters using different mapping techniques and processor designs.

\subsection{Mapping to single-core devices}

\begin{figure}[H]
	\centering
	\subfigure[] {\includegraphics[width=\linewidth]{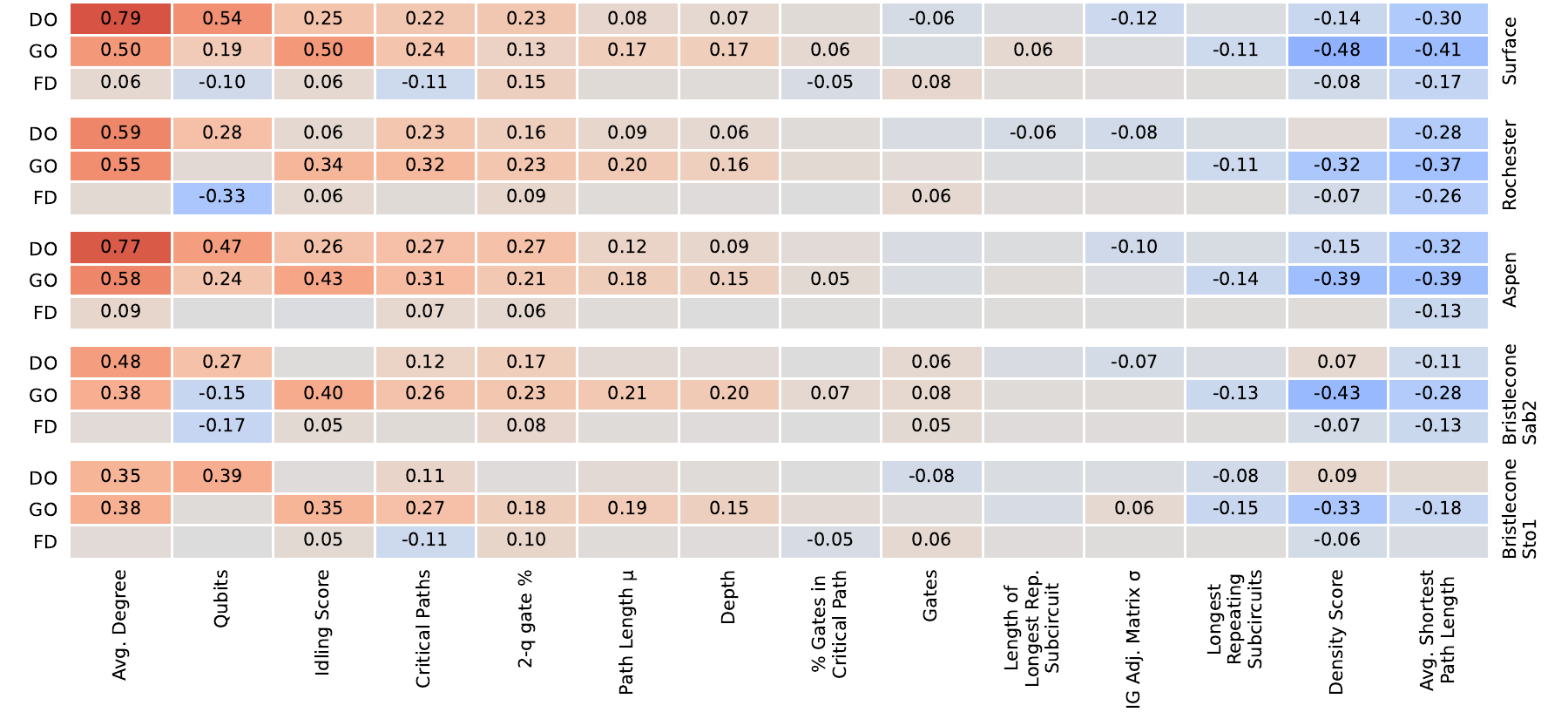}
    \label{fig:single_core_all_params}} \\
    \subfigure[] {\includegraphics[width=0.45\linewidth]{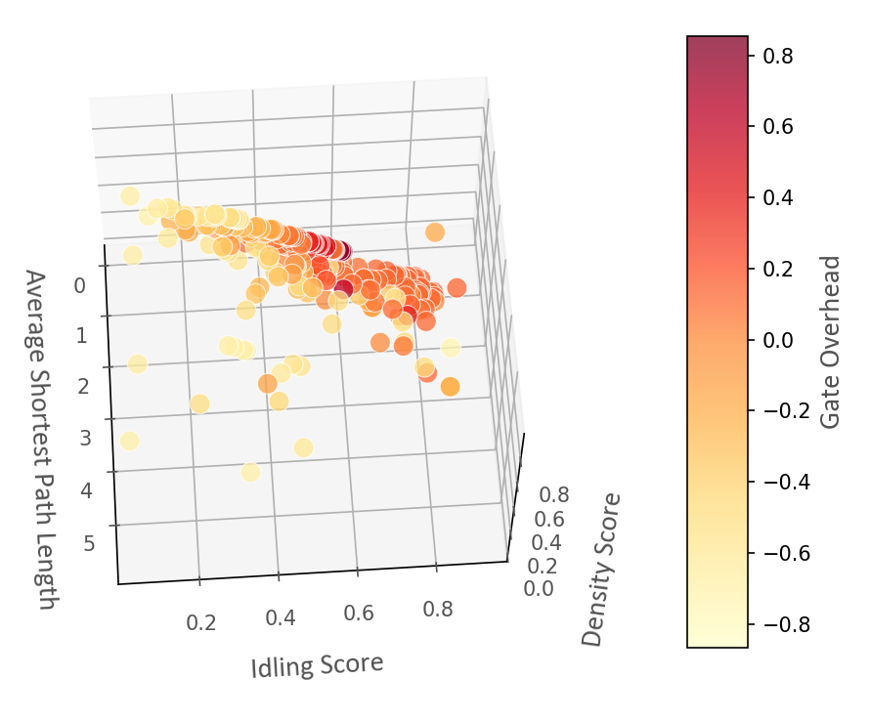}
    \label{fig:single_core_selected params1}}
    \subfigure[] {\includegraphics[width=0.46\linewidth]{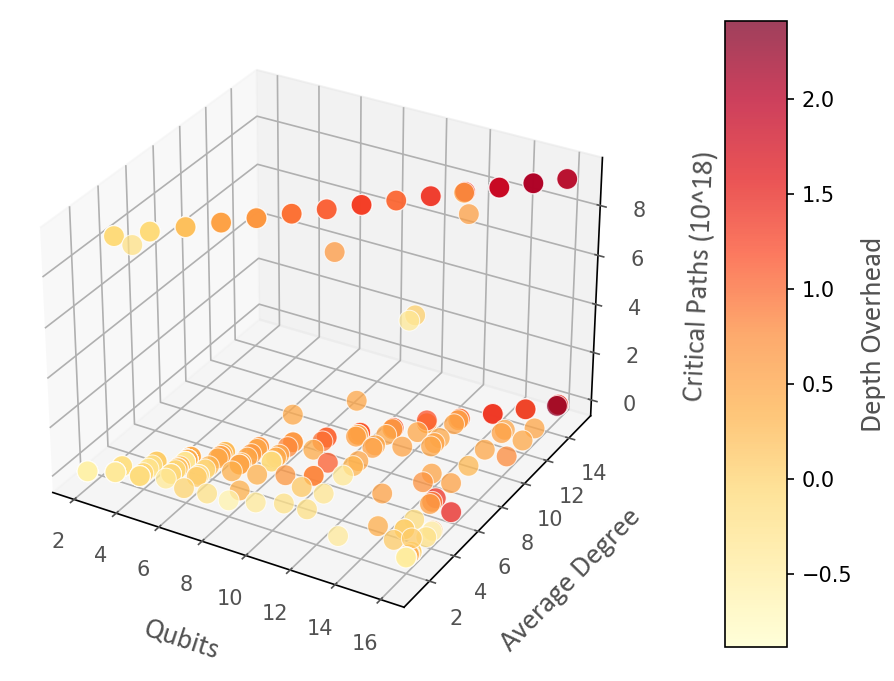}
    \label{fig:single_core_selected params2}}
    \caption{a) Correlation of the gate overhead, depth overhead and fidelity decrease compilation performance metrics with circuit parameters while using 4 different device configurations: Surface 17, IBM Rochester, Rigetti Aspen and Google Bristlecone for Sabre2 mapper, and Bristlecone device for Stochastic1 mapper. Metrics are ordered from the most positively correlated to the most negatively correlated one (red to blue). b) Three parameters with high performance correlation for the Surface 17 topology: idling score, density score and avg. shortest path length and their gate overhead. c) Three parameters with high performance correlation for the Aspen-16 topology: average degree, number of critical paths and number of qubits and their depth overhead.}
\label{fig:single_core_correlation}
\end{figure}

We begin our analysis by examining the correlation between the circuit parameters introduced in Sec. \ref{Sec3} and the three quantum circuit mapping performance metrics across four different single-core architectures: Surface 17, IBM Rochester, Rigetti Aspen, and Google Bristlecone (see Fig. \ref{fig:scdevices}). The correlations are visualized in Fig. \ref{fig:single_core_all_params} using a Pearson correlation map, where the correlation values range from dark red for the highest positive correlation to dark blue for the highest negative correlation. A positive correlation means that if values of one parameter increase, then so do of the other, and a negative is vice versa. White color, and correlation factors around zero indicate no significant correlation. The parameters are sorted to facilitate the identification of the most correlated ones. Notably, while some parameters are consistently important across all devices and metrics, others are not significant (e.g., the number of gates), and some vary in importance depending on the metrics and devices (e.g., the number of qubits has a high negative correlation factor only for the fidelity decrease in the IBM Rochester device). In addition to the analysis of the four devices using the same mapping technique (Sabre2), Fig. \ref{fig:single_core_all_params} also includes results for two different mapping configurations, namely Stochastic1 and Sabre2 (see Sec. \ref{Sec:4.3}), for the Google Bristlecone device. The configurations Stochastic2 and Sabre1 are not shown, as their correlation results are similar to Stochastic1 and Sabre2, respectively. This suggests that the routing technique is more significant than other optimization passes concerning different circuit parameters. For the same reason, we opted to use Sabre2 and Stochastic1 for the rest of the experiments explained in this section.

To investigate the meaning of these correlations, we plotted 3D graphs of some of the highest-correlated parameters. As examples, we selected: 1) the gate overhead metric for the Surface 17 device with parameters: average shortest path length, density score, and idling score, shown in Fig. \ref{fig:single_core_selected params1}, and 2) the depth overhead metric for the Rigetti Aspen device with circuit parameters: number of qubits, average IG degree, and number of critical paths, shown in Fig. \ref{fig:single_core_selected params2}. In Fig. \ref{fig:single_core_selected params1}, we observe that the average shortest path length and density score metrics are negatively correlated with the gate overhead, meaning that higher values of these metrics correspond to a lower gate overhead. Conversely, the idling score shows a positive correlation. This implies that a higher average shortest path length simplifies running the circuit due to a simpler interaction graph (IG), confirming our hypotheses in Sec. \ref{Sec3.1.2}. In Fig. \ref{fig:single_core_selected params2}, the average degree of the IG is shown to be highly influential for the depth overhead: a higher degree indicates a denser IG, leading to a larger depth overhead. The number of qubits and the number of critical paths also exhibit positive correlations.

\begin{figure}[h!]
	\centering
	\subfigure[] {\includegraphics[width=0.45\linewidth]{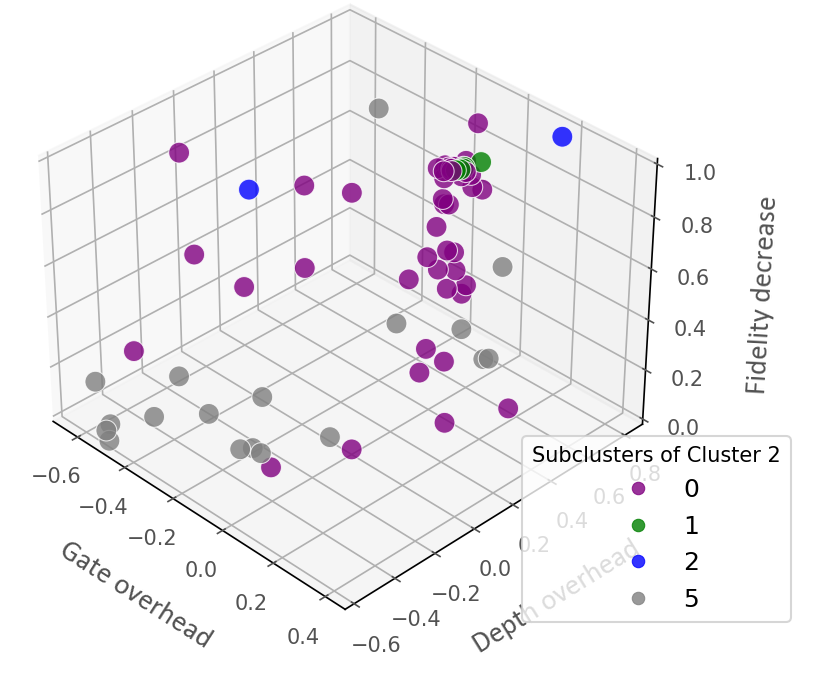}
    \label{fig:sc_clustering_sur}}
    \subfigure[] {\includegraphics[width=0.45\linewidth]{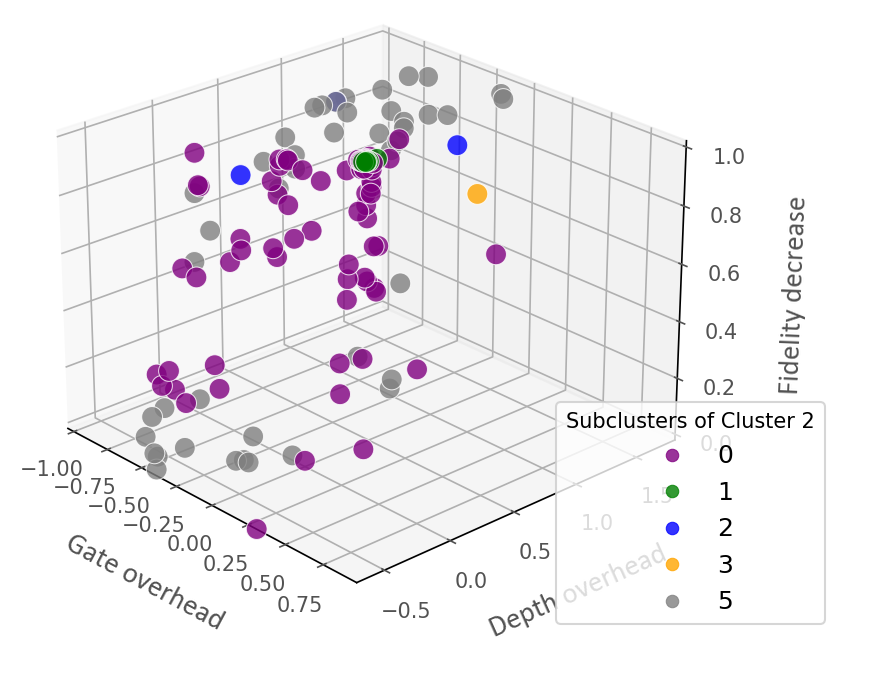}
    \label{fig:brs}}
    \subfigure[] {\includegraphics[width=0.45\linewidth]{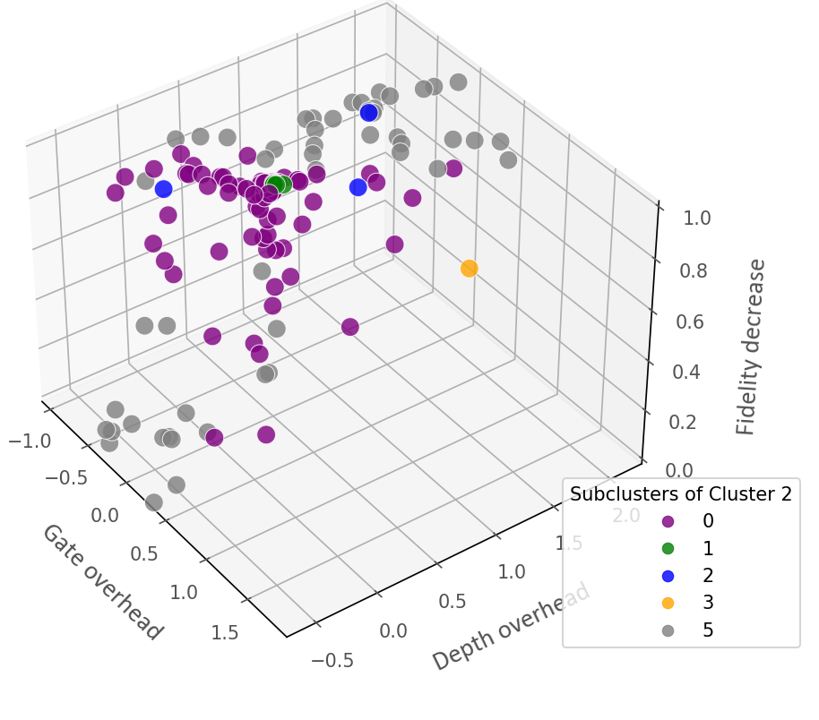}
    \label{fig:sc_clustering_brs_sto1}}
    \caption{Relation of circuit families (sub-clusters) of cluster 2 with their compilation metrics when targeting: a) Surface 17, b) Bristlecone for Sabre2 mapper, and c) Bristlecone for Stochastic1 mapper.}
\label{fig:single_core_clustering}
\end{figure}

In the next part of our work, we created ``circuit families'' containing circuits of similar structure based on extracted features. The two-level clustering approach used to achieve this is explained in Sec. \ref{sec:3.2}. Fig. \ref{fig:single_core_clustering} presents examples of these created families and their results concerning the three compilation metrics. Figs. \ref{fig:sc_clustering_sur}, \ref{fig:brs} and \ref{fig:sc_clustering_brs_sto1} illustrate performance differences when targeting Surface 17, Bristlecone using Sabre2 mapper, and Bristlecone using the Stochastic1 mapper, respectively. Notably, the group of circuits marked in purple (Group 0) performed better on average with Bristlecone than on Surface 17, exhibiting approximately 20\% fewer gates. Conversely, the group noted in grey (Group 5) performed better on Surface 17, particularly in terms of fidelity decrease, which remained below 40\% except for one outlier.  In contrast, for the Bristlecone device, the majority of benchmarks for this group ranged between 60\% and 100\%. The Stochastic1 mapper generally performed worse than Sabre2, as expected due to its simplicity. For example, Group 3 (yellow) reached 150\% in gate overhead, and Group 5 reached 200\% in depth overhead, whereas these metrics for Sabre2 were 75\% and 150\%, respectively. However, Group 0 did not show significant performance differences between the two mappers, with benchmarks performing within the same ranges (mostly between  $\sim$-50\% and $\sim$50\%) regarding gate and depth overhead. In this case, it would be preferable to use the simpler and faster Stochastic1 method. It is important to note that the majority of circuits in Group 0 are real-algorithm-based circuits, including many instances of Grover's algorithm \cite{nielsen2002quantum}, Cuccaro adder \cite{article_cuccaro_adder} and RevLib \cite{wille2008revlib} algorithms. On the other hand, most circuits in Group 5 are QUEKO circuits \cite{Queko}. This analysis provides guidance on which types of circuits work better with specific device/mapper configurations.

The Aspen and Rochester device configurations exhibited patterns very similar to Surface 17 and Bristlecone, respectively, for these particular circuit families. Therefore, they were not showcased in this paper. For the rest of the results, refer to  \ref{app:4}.

\subsection{From single-core to multi-core architectures}

\begin{figure}[h!]
	\centering
	\subfigure[]  {\includegraphics[width=\linewidth]{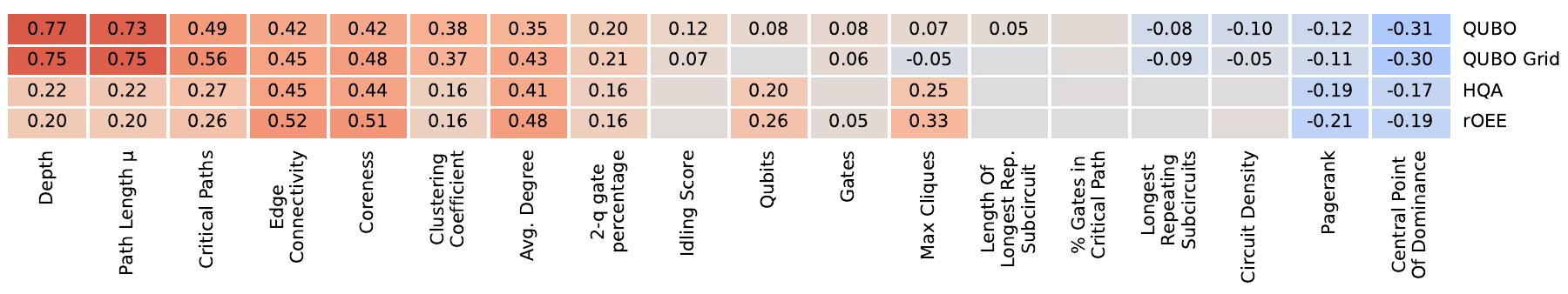}
    \label{fig:mc_all_params}} \\
    \subfigure[]   {\includegraphics[width=0.5\linewidth]{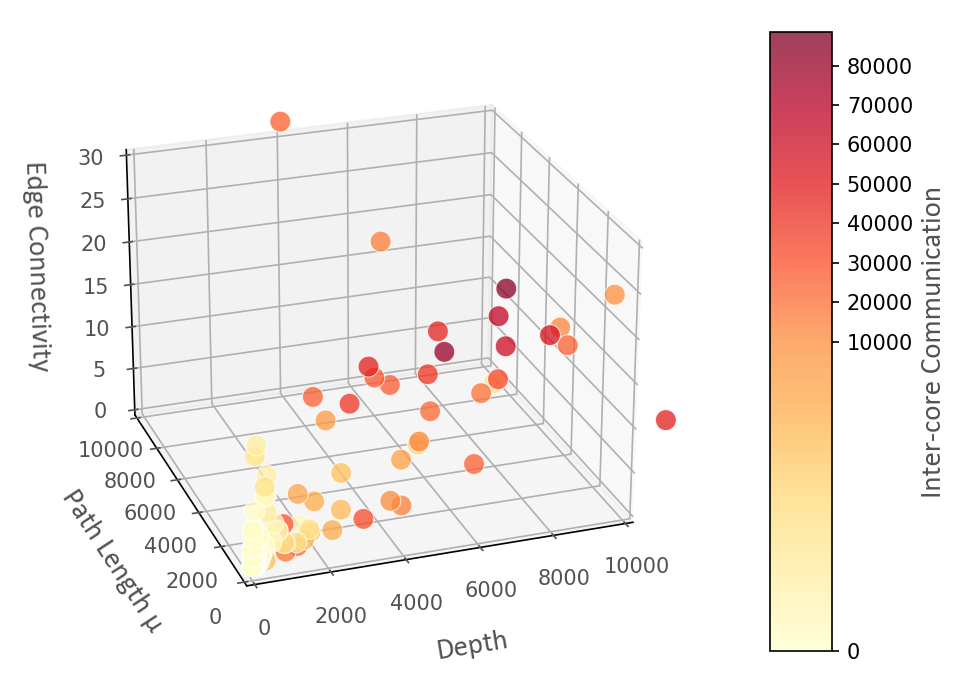}
    \label{fig:mc_selected_params}}
    \caption{a) Correlation of the number of inter-core communications performance metric with circuit parameters while using 3 different mappers: QUBO, HQA, and rOEE for all-to-all connected cores, and QUBO mapper for grid-like core connectivity. Metrics are ordered from the most positively correlated to the most negatively correlated one. b) Three highly correlated parameters for the QUBO mapper and the all-to-all connected topology: edge connectivity, mean of GDG path length and circuit depth, and their number of inter-core moves.}
 	\label{fig:multi_core_correlation}
\end{figure}

Our analysis continues by examining the correlations between the number of inter-core communications and circuit parameters using three different mapping configurations for all-to-all connected multi-core devices: QUBO, HQA, and rOEE, as introduced in Sec.\ref{Sec4} (see Fig. \ref{fig:multi_core_correlation}). Notably, the importance of circuit parameters differs among the mapping techniques. For instance, the number of qubits and maximal cliques is positively correlated with the performance when using HQA, but not with other techniques. Circuit depth and GDG path length mean exhibit a high correlation score (approximately 0.75) with QUBO, while the scores are only 0.2 and 0.4 for the other two mappers. Edge connectivity is not influential when using rOEE, but it is for the other two mappers. Additionally, we explored two different core topologies with the QUBO mapper: all-to-all and grid. Generally, the parameter correlations between these topologies are similar, though some differences exist. For example, the average degree circuit parameter is significantly more important for grid connectivity due to the less connected device connectivity graph, emphasizing the importance of IG connectivity of the circuit, as expected in Sec. \ref{Sec3}. We anticipate a larger difference when also reducing the connectivity within the cores, an experiment we leave for future work.

Fig. \ref{fig:mc_selected_params} shows an example of the correlation between parameters: edge connectivity, GDG path length mean, and circuit depth, with the number of inter-core communications when using the QUBO mapper. A trend is observed where overhead increases with higher metrics values, confirming the high positive correlation shown by their correlation factors in Fig. \ref{fig:mc_all_params}.

\begin{figure}[H]
	\centering
	\subfigure[] {\includegraphics[width=0.7\linewidth]{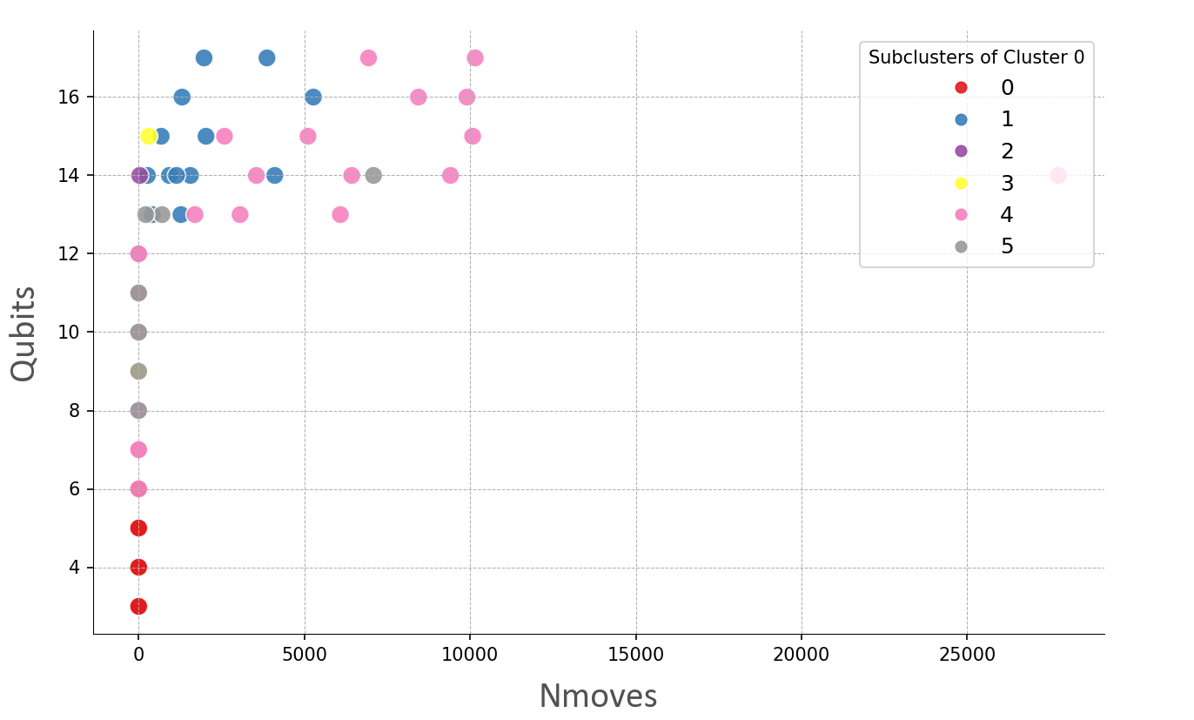}
    \label{fig:mc_clustering_hqa}}
    \subfigure[] {\includegraphics[width=0.7\linewidth]{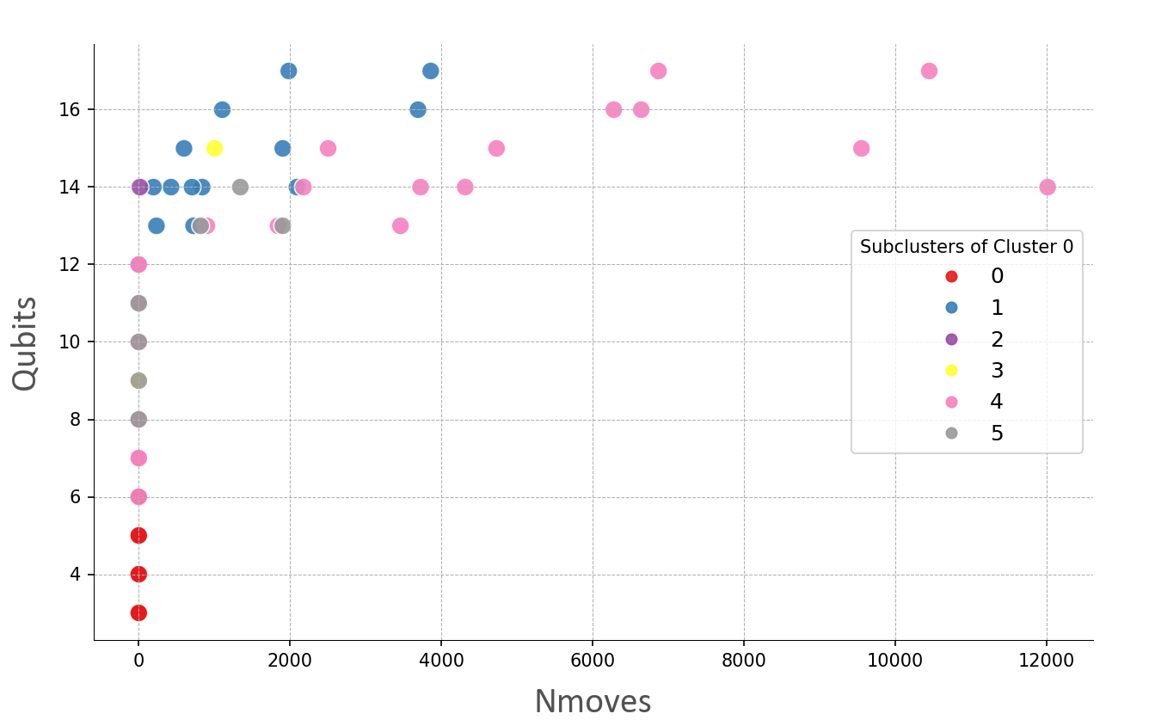}
    \label{fig:mc_clustering_roee}}
    \caption{Relation of circuit families (sub-clusters) of cluster 0 with the number of inter-core communications when using all-to-all core connectivity and a) HQA mapper and b) rOEE mapper.}
\label{fig:multi_core_clustering}
\end{figure}

Similar to single-core devices, we also analyze circuit families and their performance patterns for modular architecture. Fig. \ref{fig:multi_core_clustering} shows the groups of circuits that are part of size cluster 0 and their number of moves (inter-core communications) for the two mappers, HQA (a) and rOEE (b). Overall, the HQA mapper performed better, scaling up to $10,000$ moves (with only one outlier with $25000$) compared to $12,000$ for rOEE. rOEE reached that number only for group 4 (pink), making HQA a better choice for this group. Group 0 performed particularly well with both mappers, with all benchmarks having up to 100 moves. 
Group 4 showcased the worst performance for HQA, reaching up to $10,000$ moves, but still significantly better than rOEE. Groups 1 and 5 (blue and grey, respectively), on the other hand, showed, on average, a better performance with rOEE, where group 5 outperformed most other groups, despite the higher qubit count. Note that here we do not showcase QUBO results due to lower amount of successfully compiled instances, which would lead to unfair comparison. The results are however available at location shown in \ref{app:4}. 


The main insights are as follows:

\begin{itemize}
    \item There is an evident correlation between the structural parameters of the circuits and their performance, which varies with different compilation and device configurations. This confirms that adapting the quantum system layers to the circuit properties and employing hardware-software co-design is key to successful quantum circuit execution.
    \item It is possible to make appropriate compilation/device choices per circuit family instead of per single circuit, as there are preferred combinations of device and compilation for each group of similar circuits.
    \item IG parameters are most important for less connected device topologies.
    \item Idling and density scores are always of high importance for single-core devices.
    \item GDG-related parameters and circuit depth are the most relevant metrics for modular architectures, and they are more significant than for single-core architectures.
    \item Contrary to our expectations, standard parameters like the number of gates, qubits, and the percentage of two-qubit gates, as well as repetitive sub-circuit metrics, are not that significant for circuit success rates.
\end{itemize}

\subsection{Evaluating clusters based on circuit origin}

The final part of this section showcases the types of clusters created based on the origin of the circuits. We aim to validate that circuits within the same families exhibit similarities not only in their extracted structural parameters but also in their origin. For instance, it stands to reason that randomly generated circuits would share similar structural parameters. This validation would also support the comprehensiveness of the parameter set we used to describe the circuits. 

\begin{figure}[H]
	\centering
	\includegraphics[width=0.6 \linewidth]{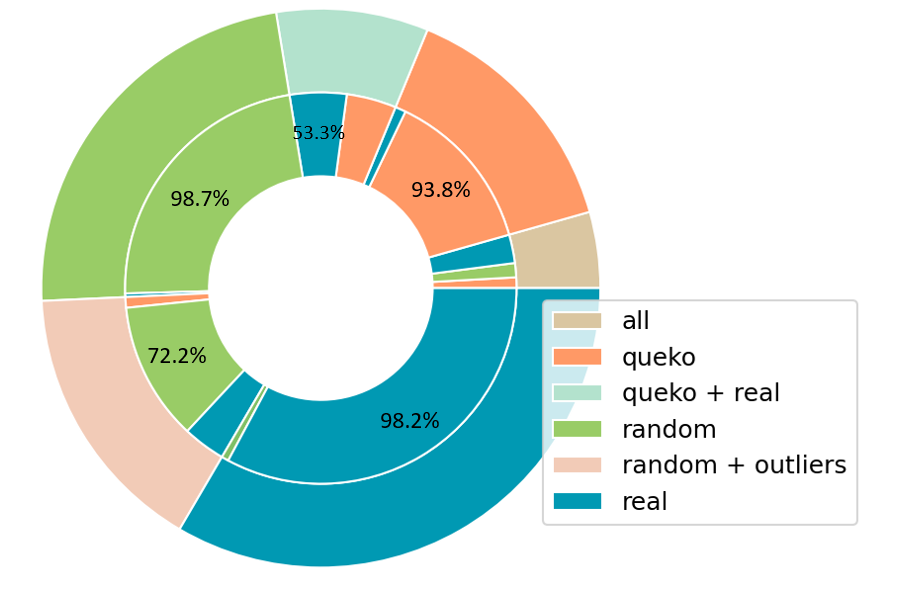}
	\caption{The distribution of the extracted circuit families based on multi-core-architectures circuit parameters per circuit type:  Each segment in the outer circle represents groups characterized by the same types of circuits (e.g., the blue segment encompasses circuits based on real algorithms). The inner circle displays the quantity of each circuit type within the respective outer circle segment.}
\label{fig:cluster_type_dostribution}
\end{figure}

For this purpose, we labeled three groups: real algorithms (based on actual quantum algorithms like Grover's or arithmetic circuits), randomly generated circuits, and QUEKO circuits (synthetic circuits with predefined depth and gate count)\cite{Queko}. The circuits were sourced from \cite{qbench}. Fig. \ref{fig:cluster_type_dostribution} shows the distribution of the existing sub-circuits after a two-level clustering for the modular-architecture-related metrics. We observe that circuits with the same origin generally belong to the same group, with a few outliers. The exceptions are a very small percentage of mixed benchmarks (noted in light brown) and sub-clusters containing both QUEKO and real benchmarks (in mint-green). Since QUEKO circuits aim to mimic realistic behavior more closely than classical random circuits \cite{Queko}, the existence of these mixed groups is expected. The mixed benchmarks group comprises those circuits that are structurally unique compared to other circuits of the same type. This figure confirms that our defined circuit parameters effectively identify structural similarities between circuits, especially highlighting the distinction of randomly generated circuits (green) compared to the other two types that contain logical patterns and oracles.

The complete set of results (for all benchmarks, clusters, compilers and devices) can be found at \noindent\href{https://github.com/QML-Group/QuantumCircuitProfiling}{https://github.com/QML-Group/QuantumCircuitProfiling}.

\section{Conclusion}
\label{sec:conclusion}

To advance the development of future quantum computers and increase the probability of successfully executing algorithms that are currently unfeasible on classical computers, it is crucial to understand their structure. Quantum circuits (i.e., executable versions of quantum algorithms) will yield the highest success rates when run on a quantum system tailored for it. In classical computing, we see two trends that could also shape the future of quantum computers: a) application-specific devices optimized for particular tasks, and b) multi-core computation as a solution for current scalability problems.

In this paper, we explored these aspects within the quantum computing field by introducing the most comprehensive characterization of quantum circuit structures to date. This characterization addresses all facets of quantum circuits: circuit size, qubit interactions, gate density and dependencies, and repetitive patterns. For this purpose, we defined 15 and 19 circuit features of importance for single-core and multi-core quantum devices, respectively. We made special emphasis on modular (i.e., multi-core) architectures, which represent the future of quantum computing and a solution to the scalability challenge of current NISQ devices. Multi-core architecture and related compilation techniques function differently than current quantum systems, because the focus is primarily on reducing the expensive communication between cores. Consequently, extra circuit parameters should be considered for multi-core architectures.

The extracted parameters were used to create families of similarly structured circuits by utilizing the k-means clustering technique. Identifying circuit families simplifies the development of high-performance architectures (by optimizing for families instead of individual circuits), helps to create a representative set of circuits (using cluster representatives), and aids in approximating the performance of future circuits without running them. We also analyzed the types of circuits within the clusters and found that similarly created circuits (real-algorithms, QUEKO, uniformly random) are mostly grouped together. This confirms that our characterization can comprehensively describe a circuit and identify structural similarities between different ones.

Furthermore, we conducted experiments with six different devices and seven mapping configurations to showcase varying correlation levels and performance patterns in terms of gate overhead, depth overhead, fidelity decrease, and number of inter-core communications metrics. 
These experiments revealed which parameters influence circuit success rates the most, and how these correlations change with different devices or compilers. We could also see that some mappers scale better with the same circuit families than others (e.g., HQA vs. rOEE mapper). Therefore, the patterns observed in clustered circuits can help identify the best existing mapper and device to use, particularly for higher numbers of qubits.

The proposed method and current findings can aid in the development of quantum systems by incorporating information about circuit structure and providing deeper insights into the variability of outcomes when executing different quantum circuits. Structural parameters of circuits can also be used to predict fidelity decrease, gate overhead, and depth overhead for specific processors and compilation techniques without executing them on actual devices. This can facilitate the co-design of quantum compilers, processors, and applications, ultimately contributing to the development of high-performance application-specific quantum systems. An in-depth analysis of quantum circuits used for benchmarking quantum computing systems is a systematic approach that directly contributes to assessing current devices and lays the foundation for designing future scalable architectures.

\section{Acknowledgements}

 SF and MB acknowledge funding of Intel Corporation. CGA acknowledges funding from the Spanish Ministry of Science, Innovation and Universities through the Beatriz Galindo program 2020 (BG20-00023) and from project PCI2022-133004 funded by MCIN/AEI/10.13039/501100011033 and by the European Union NextGenerationEU/PRTR. CGA, EA and SA also acknowledge support from the EU, grant HORIZON-EIC-2022-PATHFINDEROPEN-01-101099697. Finally, SA acknowledges support from the EU, grant ERC-StG-2021-WINC-101042080.


\section{References}

\bibliographystyle{unsrt}
\bibliography{mybibliography}

\appendix

\section{Selected quantum benchmarks}
\label{app:1}

\fontsize{6pt}{7pt}\selectfont
\begin{longtable}{|l|l|l|l|}
    \caption{Benchmarks used for the experiments.}\\
    \hline
    \textbf{Benchmark} & \textbf{Qubits} & \textbf{Gates} & Two-qubit gate percentage \\ \hline
    \endfirsthead 
    \multicolumn{4}{c}%
    {{\bfseries \tablename\ \thetable{} -- continued from previous page}} \\
    \hline
    Benchmark & Qubits & Gates & Two-qubit gate percentage \\ \hline
    \endhead
    \hline \multicolumn{4}{|r|}{{Continued on next page}} \\ \hline
    \endfoot 
    \hline
    \endlastfoot
        0410184\_169 & 14 & 211 & 0.492890995 \\ \hline
        15\_enc & 15 & 264 & 0.522727273 \\ \hline
        16QBT\_100CYC\_QSE\_1 & 16 & 1400 & 0.327142857 \\ \hline
        16QBT\_10CYC\_TFL\_1 & 16 & 1473 & 0.330617787 \\ \hline
        16QBT\_15CYC\_TFL\_2 & 16 & 1582 & 0.335651075 \\ \hline
        16QBT\_20CYC\_TFL\_3 & 16 & 1727 & 0.341053851 \\ \hline
        16QBT\_35CYC\_TFL\_6 & 16 & 1980 & 0.348484848 \\ \hline
        16QBT\_40CYC\_TFL\_7 & 16 & 2269 & 0.355222565 \\ \hline
        16QBT\_500CYC\_QSE\_3 & 16 & 7949 & 0.302679582 \\ \hline
        16QBT\_700CYC\_QSE\_5 & 16 & 15901 & 0.292182882 \\ \hline
        16QBT\_900CYC\_QSE\_7 & 16 & 26125 & 0.288076555 \\ \hline
        20QBT\_100CYC\_QSE\_8 & 20 & 27545 & 0.287747323 \\ \hline
        20QBT\_400CYC\_QSE\_8 & 20 & 33225 & 0.286711813 \\ \hline
        20QBT\_45CYC\_0D1\_1D2\_0 & 15 & 33270 & 0.287676586 \\ \hline
        20QBT\_45CYC\_0D1\_1D2\_5 & 13 & 33315 & 0.288638751 \\ \hline
        20QBT\_45CYC\_0D1\_2D2\_0 & 20 & 33405 & 0.290555306 \\ \hline
        20QBT\_45CYC\_0D1\_2D2\_5 & 20 & 33495 & 0.292461561 \\ \hline
        20QBT\_45CYC\_0D1\_2D2\_9 & 20 & 33585 & 0.2943576 \\ \hline
        20QBT\_45CYC\_0D1\_3D2\_0 & 20 & 33720 & 0.297182681 \\ \hline
        20QBT\_45CYC\_0D1\_3D2\_5 & 20 & 33855 & 0.299985231 \\ \hline
        20QBT\_45CYC\_0D1\_3D2\_9 & 20 & 33990 & 0.302765519 \\ \hline
        20QBT\_45CYC\_0D1\_4D2\_0 & 20 & 34170 & 0.306438396 \\ \hline
        20QBT\_45CYC\_0D1\_5D2\_1 & 20 & 34395 & 0.310975432 \\ \hline
        20QBT\_45CYC\_0D1\_6D2\_2 & 20 & 34665 & 0.316342132 \\ \hline
        20QBT\_45CYC\_0D1\_7D2\_3 & 20 & 34980 & 0.322498571 \\ \hline
        20QBT\_45CYC\_0D1\_8D2\_4 & 20 & 35340 & 0.329400113 \\ \hline
        20QBT\_45CYC\_1D1\_1D2\_5 & 20 & 35475 & 0.329415081 \\ \hline
        20QBT\_45CYC\_1D1\_2D2\_6 & 20 & 35655 & 0.330276259 \\ \hline
        20QBT\_45CYC\_1D1\_3D2\_7 & 20 & 35880 & 0.33196767 \\ \hline
        20QBT\_45CYC\_1D1\_4D2\_8 & 20 & 36150 & 0.334467497 \\ \hline
        20QBT\_45CYC\_1D1\_5D2\_9 & 20 & 36465 & 0.337748526 \\ \hline
        20QBT\_45CYC\_1D1\_6D2\_0 & 20 & 36825 & 0.341778683 \\ \hline
        20QBT\_45CYC\_1D1\_7D2\_1 & 20 & 37230 & 0.346521622 \\ \hline
        20QBT\_45CYC\_2D1\_2D2\_3 & 20 & 37500 & 0.346426667 \\ \hline
        20QBT\_45CYC\_2D1\_3D2\_4 & 20 & 37815 & 0.347110935 \\ \hline
        20QBT\_45CYC\_2D1\_4D2\_5 & 20 & 38175 & 0.348552718 \\ \hline
        20QBT\_45CYC\_2D1\_5D2\_6 & 20 & 38580 & 0.350725765 \\ \hline
        20QBT\_45CYC\_2D1\_6D2\_7 & 20 & 39030 & 0.353599795 \\ \hline
        20QBT\_45CYC\_3D1\_1D2\_8 & 17 & 39345 & 0.351912568 \\ \hline
        20QBT\_45CYC\_3D1\_2D2\_9 & 20 & 39705 & 0.35098854 \\ \hline
        20QBT\_45CYC\_3D1\_3D2\_0 & 20 & 40110 & 0.350810272 \\ \hline
        20QBT\_45CYC\_3D1\_4D2\_1 & 20 & 40560 & 0.351356016 \\ \hline
        20QBT\_45CYC\_3D1\_5D2\_2 & 20 & 41055 & 0.352600171 \\ \hline
        20QBT\_45CYC\_4D1\_1D2\_3 & 19 & 41460 & 0.350241196 \\ \hline
        20QBT\_45CYC\_4D1\_2D2\_4 & 20 & 41910 & 0.348628012 \\ \hline
        20QBT\_45CYC\_4D1\_3D2\_5 & 20 & 42405 & 0.347742012 \\ \hline
        20QBT\_45CYC\_4D1\_4D2\_6 & 20 & 42945 & 0.347560834 \\ \hline
        20QBT\_45CYC\_5D1\_1D2\_7 & 20 & 43440 & 0.34463628 \\ \hline
        20QBT\_45CYC\_5D1\_2D2\_8 & 20 & 43980 & 0.342451114 \\ \hline
        20QBT\_45CYC\_5D1\_3D2\_9 & 20 & 44565 & 0.340985078 \\ \hline
        20QBT\_45CYC\_6D1\_2D2\_1 & 20 & 45195 & 0.338223255 \\ \hline
        20QBT\_500CYC\_QSE\_7 & 20 & 52295 & 0.330547854 \\ \hline
        20QBT\_800CYC\_QSE\_4 & 20 & 63655 & 0.321828607 \\ \hline
        20QBT\_900CYC\_QSE\_3 & 20 & 76435 & 0.31511742 \\ \hline
        3\_17\_13 & 3 & 76471 & 0.31519138 \\ \hline
        4gt10-v1\_81 & 5 & 76619 & 0.31544395 \\ \hline
        4gt11\_82 & 5 & 76646 & 0.315567675 \\ \hline
        4gt12-v0\_87 & 6 & 76893 & 0.31601056 \\ \hline
        4gt13\_92 & 5 & 76959 & 0.316129368 \\ \hline
        4gt4-v0\_72 & 6 & 77217 & 0.316536514 \\ \hline
        4gt5\_75 & 5 & 77300 & 0.316688228 \\ \hline
        4mod5-bdd\_287 & 7 & 77370 & 0.316802378 \\ \hline
        4mod7-v0\_94 & 5 & 77532 & 0.317069081 \\ \hline
        4\_49\_16 & 5 & 77749 & 0.317457459 \\ \hline
        53QBT\_700CYC\_QSE\_3 & 53 & 104091 & 0.308412831 \\ \hline
        54QBT\_10CYC\_QSE\_9 & 54 & 104475 & 0.308312994 \\ \hline
        54QBT\_15CYC\_QSE\_8 & 54 & 105051 & 0.308164606 \\ \hline
        54QBT\_200CYC\_QSE\_7 & 54 & 112719 & 0.306363612 \\ \hline
        54QBT\_25CYC\_QSE\_6 & 54 & 113678 & 0.306154225 \\ \hline
        54QBT\_35CYC\_QSE\_4 & 54 & 115020 & 0.305868545 \\ \hline
        54QBT\_40CYC\_QSE\_3 & 54 & 116554 & 0.305549359 \\ \hline
        54QBT\_500CYC\_QSE\_0 & 54 & 135724 & 0.302179423 \\ \hline
        54QBT\_800CYC\_QSE\_7 & 54 & 166396 & 0.298402606 \\ \hline
        9symml\_195 & 11 & 201277 & 0.322366689 \\ \hline
        adder\_n10 & 10 & 201571 & 0.32221897 \\ \hline
        adder\_n4 & 4 & 201594 & 0.322231812 \\ \hline
        adr4\_197 & 13 & 205033 & 0.324133188 \\ \hline
        aj-e11\_165 & 5 & 205184 & 0.324230934 \\ \hline
        alu-bdd\_288 & 7 & 205268 & 0.324283376 \\ \hline
        alu-v0\_26 & 5 & 205352 & 0.324335775 \\ \hline
        alu-v1\_28 & 5 & 205389 & 0.324364985 \\ \hline
        alu-v2\_30 & 6 & 205893 & 0.324654068 \\ \hline
        alu-v2\_31 & 5 & 206344 & 0.324904044 \\ \hline
        alu-v3\_34 & 5 & 206396 & 0.324938468 \\ \hline
        alu-v4\_36 & 5 & 206511 & 0.325004479 \\ \hline
        basis\_change\_n3 & 3 & 206590 & 0.324928603 \\ \hline
        basis\_trotter\_n4 & 4 & 208528 & 0.323836607 \\ \hline
        bell\_n4 & 4 & 208577 & 0.32379409 \\ \hline
    benstein\_vazirani\_1b\_secret\_1 & 2 & 208584 & 0.323792812 \\ \hline
    benstein\_vazirani\_28b\_secret\_64 & 2 & 208645 & 0.323707733 \\ \hline
        bigadder\_n18 & 18 & 209231 & 0.323422437 \\ \hline
        bv\_n14 & 14 & 209272 & 0.323421193 \\ \hline
        bv\_n19 & 19 & 209328 & 0.32342066 \\ \hline
        C17\_204 & 7 & 209795 & 0.323677876 \\ \hline
        cat\_state\_n4 & 4 & 209799 & 0.323686004 \\ \hline
        cm82a\_208 & 8 & 210449 & 0.324031 \\ \hline
        cnt3-5\_179 & 16 & 210624 & 0.324165337 \\ \hline
        co14\_215 & 15 & 228560 & 0.333028526 \\ \hline
        con1\_216 & 9 & 229514 & 0.333452426 \\ \hline
        cuccaroAdder\_10b & 22 & 230235 & 0.333107477 \\ \hline
        CuccaroAdder\_1b & 4 & 230308 & 0.333075707 \\ \hline
        cuccaroadder\_q128 & 127 & 232979 & 0.333785448 \\ \hline
        cuccaroadder\_q16 & 15 & 233258 & 0.333827779 \\ \hline
        cuccaroadder\_q32 & 31 & 233897 & 0.333937588 \\ \hline
        cuccaroadder\_q64 & 63 & 235256 & 0.334180637 \\ \hline
        cuccaroMultiplier\_1b & 5 & 235432 & 0.334066737 \\ \hline
        cycle10\_2\_110 & 12 & 241482 & 0.336662774 \\ \hline
        dc1\_220 & 11 & 243396 & 0.337437756 \\ \hline
        dc2\_222 & 15 & 252858 & 0.341147996 \\ \hline
        decod24-v1\_41 & 5 & 252943 & 0.341183587 \\ \hline
        deutsch\_n2 & 2 & 252948 & 0.341180796 \\ \hline
        dist\_223 & 13 & 290994 & 0.353701451 \\ \hline
        dnn\_n16 & 16 & 294034 & 0.351350524 \\ \hline
        dnn\_n2 & 2 & 294372 & 0.351089778 \\ \hline
        dnn\_n8 & 8 & 295892 & 0.349935111 \\ \hline
        error\_correctiond3\_n5 & 5 & 296005 & 0.349967061 \\ \hline
        ex-1\_166 & 3 & 296024 & 0.349975002 \\ \hline
        ex3\_229 & 6 & 296427 & 0.350089567 \\ \hline
        f2\_232 & 8 & 297633 & 0.350434932 \\ \hline
        fredkin\_n3 & 3 & 297652 & 0.350439439 \\ \hline
        ghz\_q128 & 127 & 297779 & 0.350713113 \\ \hline
        ghz\_q16 & 15 & 297810 & 0.350723616 \\ \hline
        ghz\_q32 & 31 & 297841 & 0.350787836 \\ \hline
        ghz\_q64 & 64 & 297905 & 0.350923952 \\ \hline
        graycode6\_47 & 6 & 297910 & 0.350934846 \\ \hline
        grover\_n2 & 2 & 297926 & 0.350922712 \\ \hline
        grover\_n3 & 3 & 298015 & 0.350858178 \\ \hline
        grover\_operator\_8 & 8 & 298506 & 0.351554073 \\ \hline
        grover\_q128\_1 & 127 & 303497 & 0.350675624 \\ \hline
        grover\_q16\_1 & 15 & 304008 & 0.350559854 \\ \hline
        grover\_q32\_1 & 31 & 305159 & 0.350338676 \\ \hline
        grover\_q64 & 63 & 307590 & 0.349910595 \\ \hline
        ham7\_104 & 7 & 307910 & 0.350030853 \\ \hline
        hwb7\_59 & 8 & 332289 & 0.356493895 \\ \hline
        inc\_237 & 16 & 342908 & 0.358973836 \\ \hline
        ising\_model\_13 & 13 & 343541 & 0.358661703 \\ \hline
        iswap\_n2 & 2 & 343550 & 0.358658128 \\ \hline
        life\_238 & 11 & 365995 & 0.363439391 \\ \hline
        linearsolver\_n3 & 3 & 366018 & 0.363427482 \\ \hline
        lpn\_n5 & 3 & 366029 & 0.363422024 \\ \hline
        majority\_239 & 7 & 366641 & 0.36354363 \\ \hline
        max46\_240 & 10 & 393767 & 0.368578372 \\ \hline
        miller\_11 & 3 & 393817 & 0.36858998 \\ \hline
        mini-alu\_167 & 5 & 394105 & 0.368640337 \\ \hline
        misex1\_241 & 15 & 398918 & 0.369456881 \\ \hline
        mlp4\_245 & 16 & 417770 & 0.372489647 \\ \hline
        mod10\_171 & 5 & 418014 & 0.372530585 \\ \hline
        mod5adder\_127 & 6 & 418569 & 0.372607623 \\ \hline
        mod5d2\_64 & 5 & 418622 & 0.372620168 \\ \hline
        multipler\_n15 & 15 & 419880 & 0.372089645 \\ \hline
        multiply\_n13 & 11 & 420020 & 0.372027522 \\ \hline
        one-two-three-v1\_99 & 5 & 420152 & 0.372051067 \\ \hline
        plus63mod4096\_163 & 13 & 548896 & 0.387408544 \\ \hline
        pm1\_249 & 14 & 550672 & 0.3875592 \\ \hline
        q=10\_s=19990\_2qbf=02\_1 & 10 & 570672 & 0.380798427 \\ \hline
        q=10\_s=19990\_2qbf=05\_1 & 10 & 590672 & 0.384963228 \\ \hline
        q=10\_s=2990\_2qbf=03\_1 & 10 & 593672 & 0.384581048 \\ \hline
        q=10\_s=29990\_2qbf=05\_1 & 10 & 623672 & 0.390166626 \\ \hline
        q=10\_s=39990\_2qbf=08\_1 & 10 & 663672 & 0.414793754 \\ \hline
        q=10\_s=49990\_2qbf=09\_1 & 10 & 713672 & 0.448717338 \\ \hline
        q=10\_s=50\_2qbf=096\_1 & 10 & 713732 & 0.448746869 \\ \hline
        q=10\_s=90\_2qbf=011\_1 & 9 & 713832 & 0.448698013 \\ \hline
        q=10\_s=990\_2qbf=091\_1 & 10 & 714832 & 0.449327954 \\ \hline
        q=11\_s=19989\_2qbf=01\_1 & 11 & 734832 & 0.439857001 \\ \hline
        q=11\_s=2989\_2qbf=02\_1 & 11 & 737832 & 0.438870908 \\ \hline
        q=11\_s=29989\_2qbf=03\_1 & 11 & 767832 & 0.433373446 \\ \hline
        q=11\_s=39989\_2qbf=05\_1 & 11 & 807832 & 0.436708375 \\ \hline
        q=11\_s=49989\_2qbf=08\_1 & 11 & 857832 & 0.458122336 \\ \hline
        q=11\_s=49\_2qbf=061\_1 & 11 & 857892 & 0.458126431 \\ \hline
        q=11\_s=89\_2qbf=022\_1 & 10 & 857992 & 0.458092849 \\ \hline
        q=11\_s=989\_2qbf=081\_1 & 11 & 858992 & 0.458500196 \\ \hline
        q=12\_s=19988\_2qbf=01\_1 & 12 & 878992 & 0.450404554 \\ \hline
        q=12\_s=29988\_2qbf=03\_1 & 12 & 908992 & 0.445463766 \\ \hline
        q=12\_s=49988\_2qbf=05\_1 & 12 & 958992 & 0.448407286 \\ \hline
        q=12\_s=59988\_2qbf=09\_1 & 12 & 1018992 & 0.475073406 \\ \hline
        q=12\_s=9988\_2qbf=01\_1 & 12 & 1028992 & 0.471399195 \\ \hline
        q=13\_s=19987\_2qbf=01\_1 & 13 & 1048992 & 0.464347679 \\ \hline
        q=13\_s=29987\_2qbf=02\_1 & 13 & 1078992 & 0.457022851 \\ \hline
        q=13\_s=49987\_2qbf=03\_1 & 13 & 1128992 & 0.449958901 \\ \hline
        q=13\_s=59987\_2qbf=05\_1 & 13 & 1188992 & 0.452428612 \\ \hline
        q=13\_s=9987\_2qbf=08\_1 & 13 & 1198992 & 0.455379185 \\ \hline
        q=14\_s=19986\_2qbf=09\_1 & 14 & 1218992 & 0.462717557 \\ \hline
        q=14\_s=29986\_2qbf=01\_1 & 14 & 1248992 & 0.454013316 \\ \hline
        q=14\_s=39986\_2qbf=02\_1 & 14 & 1288992 & 0.44615017 \\ \hline
        q=14\_s=49986\_2qbf=03\_1 & 14 & 1338992 & 0.440653118 \\ \hline
        q=14\_s=5986\_2qbf=03\_1 & 14 & 1344992 & 0.440031614 \\ \hline
        q=14\_s=5986\_2qbf=05\_1 & 14 & 1350992 & 0.440291282 \\ \hline
        q=14\_s=59986\_2qbf=08\_1 & 14 & 1410992 & 0.455618459 \\ \hline
        q=14\_s=986\_2qbf=051\_1 & 14 & 1411992 & 0.455654848 \\ \hline
        q=14\_s=9986\_2qbf=09\_1 & 14 & 1421992 & 0.458755745 \\ \hline
        q=15\_s=19985\_2qbf=01\_1 & 15 & 1441992 & 0.453747316 \\ \hline
        q=15\_s=29985\_2qbf=02\_1 & 15 & 1471992 & 0.448508552 \\ \hline
        q=15\_s=49985\_2qbf=03\_1 & 15 & 1521992 & 0.443598258 \\ \hline
        q=15\_s=59985\_2qbf=05\_1 & 15 & 1581992 & 0.445789865 \\ \hline
        q=15\_s=985\_2qbf=051\_1 & 15 & 1582992 & 0.445810213 \\ \hline
        q=15\_s=9985\_2qbf=08\_1 & 15 & 1592992 & 0.448015433 \\ \hline
        q=16\_s=19984\_2qbf=09\_1 & 16 & 1612992 & 0.453625312 \\ \hline
        q=16\_s=29984\_2qbf=01\_1 & 16 & 1642992 & 0.44717138 \\ \hline
        q=16\_s=49984\_2qbf=02\_1 & 16 & 1692992 & 0.439795345 \\ \hline
        q=16\_s=59984\_2qbf=03\_1 & 16 & 1752992 & 0.434988865 \\ \hline
        q=16\_s=984\_2qbf=051\_1 & 16 & 1753992 & 0.435031061 \\ \hline
        q=17\_s=19983\_2qbf=05\_1 & 17 & 1773992 & 0.435784942 \\ \hline
        q=17\_s=2983\_2qbf=08\_1 & 17 & 1776992 & 0.436393636 \\ \hline
        q=17\_s=29983\_2qbf=09\_1 & 17 & 1806992 & 0.444058413 \\ \hline
        q=17\_s=43\_2qbf=028\_1 & 12 & 1807052 & 0.44405031 \\ \hline
        q=17\_s=49983\_2qbf=01\_1 & 17 & 1857052 & 0.434774578 \\ \hline
        q=17\_s=5983\_2qbf=02\_1 & 17 & 1863052 & 0.434022239 \\ \hline
        q=17\_s=59983\_2qbf=03\_1 & 17 & 1923052 & 0.42978349 \\ \hline
        q=17\_s=983\_2qbf=031\_1 & 17 & 1924052 & 0.42970928 \\ \hline
        q=17\_s=9983\_2qbf=05\_1 & 17 & 1934052 & 0.430105292 \\ \hline
        q=3\_s=19997\_2qbf=01\_1 & 3 & 1954052 & 0.426738388 \\ \hline
        q=3\_s=2997\_2qbf=01\_1 & 3 & 1957052 & 0.42625592 \\ \hline
        q=3\_s=2997\_2qbf=02\_1 & 3 & 1960052 & 0.425914721 \\ \hline
        q=3\_s=29997\_2qbf=03\_1 & 3 & 1990052 & 0.423986408 \\ \hline
        q=3\_s=39997\_2qbf=05\_1 & 3 & 2030052 & 0.42552309 \\ \hline
        q=3\_s=57\_2qbf=011\_1 & 3 & 2030112 & 0.425511991 \\ \hline
        q=3\_s=5997\_2qbf=08\_1 & 3 & 2036112 & 0.426616021 \\ \hline
        q=3\_s=59997\_2qbf=09\_1 & 3 & 2096112 & 0.440210256 \\ \hline
        q=3\_s=97\_2qbf=01\_1 & 3 & 2096212 & 0.440194026 \\ \hline
        q=3\_s=997\_2qbf=02\_1 & 3 & 2097212 & 0.440084264 \\ \hline
        q=3\_s=9997\_2qbf=03\_1 & 3 & 2107212 & 0.439416632 \\ \hline
        q=4\_s=19996\_2qbf=02\_1 & 4 & 2127212 & 0.437199959 \\ \hline
        q=4\_s=19996\_2qbf=05\_1 & 4 & 2147212 & 0.437750441 \\ \hline
        q=4\_s=2996\_2qbf=08\_1 & 4 & 2150212 & 0.438275389 \\ \hline
        q=4\_s=29996\_2qbf=09\_1 & 4 & 2180212 & 0.444609056 \\ \hline
        q=4\_s=39996\_2qbf=01\_1 & 4 & 2220212 & 0.438430654 \\ \hline
        q=4\_s=49996\_2qbf=02\_1 & 4 & 2270212 & 0.433220774 \\ \hline
        q=4\_s=49996\_2qbf=09\_1 & 4 & 2320212 & 0.443277166 \\ \hline
        q=4\_s=56\_2qbf=032\_1 & 4 & 2320272 & 0.44327303 \\ \hline
        q=4\_s=96\_2qbf=052\_1 & 4 & 2320372 & 0.443277199 \\ \hline
        q=4\_s=996\_2qbf=03\_1 & 4 & 2321372 & 0.443202985 \\ \hline
        q=4\_s=9996\_2qbf=05\_1 & 4 & 2331376 & 0.443385794 \\ \hline
        q=5\_s=19995\_2qbf=03\_1 & 5 & 2351376 & 0.442204479 \\ \hline
        q=5\_s=19995\_2qbf=08\_1 & 5 & 2371376 & 0.445262582 \\ \hline
        q=5\_s=2995\_2qbf=09\_1 & 5 & 2374376 & 0.445835874 \\ \hline
        q=5\_s=39995\_2qbf=01\_1 & 5 & 2414376 & 0.440142298 \\ \hline
        q=5\_s=49995\_2qbf=02\_1 & 5 & 2464376 & 0.435259473 \\ \hline
        q=5\_s=55\_2qbf=087\_1 & 5 & 2464436 & 0.435269571 \\ \hline
        q=5\_s=95\_2qbf=095\_1 & 5 & 2464536 & 0.435288428 \\ \hline
        q=5\_s=9995\_2qbf=03\_1 & 5 & 2474536 & 0.434721903 \\ \hline
        q=6\_s=19994\_2qbf=05\_1 & 6 & 2494536 & 0.435235651 \\ \hline
        q=6\_s=2994\_2qbf=08\_1 & 6 & 2497536 & 0.435676202 \\ \hline
        q=6\_s=29994\_2qbf=09\_1 & 6 & 2527536 & 0.441167604 \\ \hline
        q=6\_s=49994\_2qbf=01\_1 & 6 & 2577536 & 0.434544852 \\ \hline
        q=6\_s=54\_2qbf=022\_1 & 6 & 2577596 & 0.434540168 \\ \hline
        q=6\_s=94\_2qbf=053\_1 & 6 & 2577696 & 0.434543484 \\ \hline
        q=6\_s=9994\_2qbf=02\_1 & 6 & 2587696 & 0.433640196 \\ \hline
        q=7\_s=19993\_2qbf=03\_1 & 7 & 2607696 & 0.432629417 \\ \hline
        q=7\_s=2993\_2qbf=05\_1 & 7 & 2610696 & 0.432714495 \\ \hline
        q=7\_s=2993\_2qbf=08\_1 & 7 & 2613696 & 0.433130708 \\ \hline
        q=7\_s=29993\_2qbf=08\_1 & 7 & 2643696 & 0.437292336 \\ \hline
        q=7\_s=39993\_2qbf=09\_1 & 7 & 2683696 & 0.444200088 \\ \hline
        q=7\_s=53\_2qbf=034\_1 & 7 & 2683756 & 0.44419761 \\ \hline
        q=7\_s=59993\_2qbf=09\_1 & 7 & 2743756 & 0.454141695 \\ \hline
        q=7\_s=93\_2qbf=054\_1 & 7 & 2743856 & 0.454144095 \\ \hline
        q=7\_s=993\_2qbf=081\_1 & 7 & 2744856 & 0.454264996 \\ \hline
        q=8\_s=19992\_2qbf=02\_1 & 8 & 2764856 & 0.452427902 \\ \hline
        q=8\_s=2992\_2qbf=01\_1 & 8 & 2767856 & 0.452036883 \\ \hline
        q=8\_s=2992\_2qbf=03\_1 & 8 & 2770856 & 0.451846289 \\ \hline
        q=8\_s=29992\_2qbf=05\_1 & 8 & 2800856 & 0.452341356 \\ \hline
        q=8\_s=39992\_2qbf=08\_1 & 8 & 2840856 & 0.457238593 \\ \hline
        q=8\_s=49992\_2qbf=09\_1 & 8 & 2890856 & 0.464903129 \\ \hline
        q=8\_s=52\_2qbf=104\_1 & 8 & 2890916 & 0.464911468 \\ \hline
        q=8\_s=92\_2qbf=011\_1 & 8 & 2891016 & 0.464901267 \\ \hline
        q=8\_s=992\_2qbf=081\_1 & 8 & 2892016 & 0.465010567 \\ \hline
        q=9\_s=19991\_2qbf=08\_1 & 9 & 2912016 & 0.467286924 \\ \hline
        q=9\_s=2991\_2qbf=01\_1 & 9 & 2915016 & 0.466909959 \\ \hline
        q=9\_s=51\_2qbf=012\_1 & 7 & 2915076 & 0.46690275 \\ \hline
        q=9\_s=51\_2qbf=059\_1 & 9 & 2915136 & 0.466902745 \\ \hline
        q=9\_s=91\_2qbf=088\_1 & 9 & 2915236 & 0.466914514 \\ \hline
        q=9\_s=991\_2qbf=091\_1 & 9 & 2916236 & 0.467064394 \\ \hline
        qaoaWS\_128 & 128 & 2917388 & 0.467055462 \\ \hline
        qaoaWS\_16 & 16 & 2917532 & 0.467054346 \\ \hline
        qaoaWS\_32 & 32 & 2917820 & 0.467052114 \\ \hline
        qaoaWS\_64 & 64 & 2918396 & 0.467047652 \\ \hline
        qaoa\_128 & 128 & 2931068 & 0.467823333 \\ \hline
        qaoa\_16 & 15 & 2931579 & 0.467790907 \\ \hline
        qaoa\_32 & 31 & 2932730 & 0.467721884 \\ \hline
        qaoa\_64 & 64 & 2935937 & 0.467895599 \\ \hline
        qaoa\_n6 & 6 & 2936351 & 0.46784802 \\ \hline
        qec\_en\_n5 & 5 & 2936376 & 0.467847442 \\ \hline
        qec\_sm\_n5 & 5 & 2936401 & 0.467846864 \\ \hline
        qft\_8 & 8 & 2936577 & 0.467841981 \\ \hline
        qft\_n15 & 15 & 2937117 & 0.467827465 \\ \hline
        qft\_n20 & 20 & 2938087 & 0.467802349 \\ \hline
        qft\_q128 & 128 & 2970919 & 0.468104314 \\ \hline
        qft\_q16 & 16 & 2971439 & 0.468103165 \\ \hline
        qft\_q32 & 32 & 2973503 & 0.468111853 \\ \hline
        qft\_q64 & 64 & 2981727 & 0.468172975 \\ \hline
        QuantumVolume\_128 & 128 & 3178463 & 0.454658745 \\ \hline
        QuantumVolume\_16 & 16 & 3181551 & 0.454458847 \\ \hline
        QuantumVolume\_32 & 32 & 3193871 & 0.453667665 \\ \hline
        QuantumVolume\_64 & 64 & 3243087 & 0.45057194 \\ \hline
        quantum\_volume\_8 & 8 & 3243951 & 0.450481527 \\ \hline
        quantum\_volume\_n5 & 5 & 3244289 & 0.450445691 \\ \hline
        queko\_128 & 128 & 3252482 & 0.450318557 \\ \hline
        queko\_16 & 16 & 3252611 & 0.450316684 \\ \hline
        queko\_32 & 32 & 3253124 & 0.450308688 \\ \hline
        queko\_64 & 64 & 3255173 & 0.450277143 \\ \hline
        queko\_8 & 8 & 3255205 & 0.450276403 \\ \hline
        radd\_250 & 13 & 3258418 & 0.450263594 \\ \hline
        rd32\_270 & 5 & 3258502 & 0.450263035 \\ \hline
        rd53\_311 & 13 & 3258777 & 0.450263089 \\ \hline
        rd73\_140 & 10 & 3259007 & 0.450263224 \\ \hline
        rd73\_252 & 10 & 3264328 & 0.450239682 \\ \hline
        rd84\_142 & 15 & 3264671 & 0.450239549 \\ \hline
        root\_255 & 13 & 3281830 & 0.450168656 \\ \hline
        sao2\_257 & 14 & 3320407 & 0.450017423 \\ \hline
        sat\_n11 & 11 & 3321884 & 0.449893193 \\ \hline
        seca\_n11 & 11 & 3322217 & 0.449873383 \\ \hline
        sf\_274 & 6 & 3322998 & 0.449868763 \\ \hline
        shor\_15 & 11 & 3327790 & 0.449758248 \\ \hline
        shor\_35 & 15 & 3344319 & 0.449359944 \\ \hline
        simon\_n6 & 5 & 3344401 & 0.449353113 \\ \hline
        sqn\_258 & 10 & 3354624 & 0.449312948 \\ \hline
        sqrt8\_260 & 12 & 3357633 & 0.449301636 \\ \hline
        squar5\_261 & 13 & 3359626 & 0.449293761 \\ \hline
        square\_root\_7 & 15 & 3367256 & 0.449193052 \\ \hline
        sym9\_148 & 10 & 3388760 & 0.449118852 \\ \hline
        sys6-v0\_111 & 10 & 3388975 & 0.449119276 \\ \hline
        teleportation\_n3 & 3 & 3388983 & 0.449118806 \\ \hline
        toffoli\_n3 & 3 & 3389001 & 0.449118191 \\ \hline
        urf5\_280 & 9 & 3438830 & 0.449520913 \\ \hline
        variational\_n4 & 4 & 3438884 & 0.449518507 \\ \hline
        vbeAdder\_1b & 4 & 3438954 & 0.449513428 \\ \hline
        vbeAdder\_5b & 16 & 3439584 & 0.449467726 \\ \hline
        VQEHEA1\_128 & 128 & 3443414 & 0.449336618 \\ \hline
        VQEHEA1\_16 & 16 & 3443884 & 0.44931885 \\ \hline
        VQEHEA1\_32 & 32 & 3444834 & 0.449284929 \\ \hline
        VQEHEA1\_64 & 64 & 3446744 & 0.449218741 \\ \hline
        VQEHEA2\_128 & 128 & 3450574 & 0.449088181 \\ \hline
        VQEHEA2\_16 & 16 & 3451044 & 0.449070484 \\ \hline
        VQEHEA2\_32 & 32 & 3451994 & 0.449036702 \\ \hline
        VQEHEA2\_64 & 64 & 3453904 & 0.448970788 \\ \hline
        vqe\_uccsd\_n4 & 4 & 3454124 & 0.448967669 \\ \hline
        vqe\_uccsd\_n6 & 6 & 3456406 & 0.448975612 \\ \hline
        vqe\_uccsd\_n8 & 8 & 3467214 & 0.449158892 \\ \hline
        wim\_266 & 11 & 3468200 & 0.449154316 \\ \hline
        wstate\_n3 & 3 & 3468249 & 0.449150566 \\ \hline
        xor5\_254 & 6 & 3468256 & 0.449151101 \\ \hline
        z4\_268 & 11 & 3471329 & 0.449140372 
        
    \label{tab:benchmarks}
\end{longtable}

\normalsize

\section{Generating GDG parameters}
\label{app:2}

The construction of the GDG is done by a linear scan of the circuit, adding a new node $w$ for each gate. For each newly added gate node operating on $n$ qubits $(q_i)_{i=1}^n$, the algorithm adds incoming edges $(v_1, w), \dots, (v_n, w)$ such that $v_i$ is the last gate operating on qubit $q_i$. This last information is stored in and retrieved from an array that maps qubit indices to the last gate using said qubit. When no such gate exists, the edge comes from an extra sentinel node called $\mathbf{source}$. After all gates are processed, all children-less nodes are linked to another sentinel node called the $\mathbf{sink}$.

Since a quantum gate cannot depend on itself, the resulting directed graph is always \textit{acyclic}, therefore constitutes a \textit{directed acyclic graph} (DAG). We ignore the trivial case where this DAG is not weakly connected. Any \textit{topological ordering} of the nodes gives a valid sequential order for the gates during circuit execution.

The \textit{critical path length} is easily computed with an existing library, for instance using \texttt{networkx.dag\_longest\_path\_length}. Our other metrics require however custom processing of the graph, since brute-force extraction is doomed to fail for large circuits: for instance, in some DAG families, the \textit{number of critical paths} grows exponentially with the number of circuit gates.

Our solution is to \textit{recursively} compute those metrics. The recurrence relations between a node's values and the values of its children are given in Tab. \ref{table:rec_gdg}. Our Python implementation (see \ref{app:4}) is made iterative instead of recursive using a bottom-up traversal of the GDG, where nodes are iterated in \textit{reversed} topological ordering.

\begin{table}
\small
\begin{center}
\begin{tabular}{ | p{0.2\linewidth} | p{0.7\linewidth} | }
\hline
\textbf{Metric} & \textbf{Initialisation and recurrence relation} \\
\hline
Number of gates in critical paths to $\mathbf{sink}$ ($L$) & $$L(\mathbf{sink}) = 0$$
$$L(w) = 1 + \max_i L(v_i)$$
$$\text{From that we can define an edge predicate describing}$$
$$\text{whether a given edge } (w, v_i) \text{ belongs to a critical path:}$$
$$\mathbf{CP}(w, v_i) = \top \text{ if } L(w) = 1 + L(v_i) \text{,  } \bot \text{ otherwise.}$$
\\
\hline
Number of paths to $\mathbf{sink}$ ($n$) & $$n(\mathbf{sink}) = 1$$
$$n(w) = \sum_i n(v_i)$$
\\
\hline
Number of critical paths to $\mathbf{sink}$ ($N$) & $$N(\mathbf{sink}) = 1$$
$$N(w) = \sum_{i\ | \ \mathbf{CP}(w, v_i)} N(v_i)$$ \\
\hline
Max number of two-qubit gates in critical paths to $\mathbf{sink}$ ($M$) & $$M(\mathbf{sink}) = 0$$
    $$M(w) = \begin{cases}
      1 + \max_{i\ |\ \mathbf{CP}(w, v_i)} M(v_i) & \text{if $w$ is a 2-qubit gate} \\
      \max_{i\ |\ \mathbf{CP}(w, v_i)} M(v_i) & \text{otherwise}
    \end{cases}$$ \\
\hline
Number of critical paths to $\mathbf{sink}$ with max two-qubit gates ($K$) & $$K(\mathbf{sink}) = 1$$
$$K(w) = \sum_{i\ |\ \mathbf{CP}(w, v_i) \wedge M(v_i) + 1 = M(w)} K(v_i) $$ \\
\hline

Mean length of paths to $\mathbf{sink}$ ($m$) &
$$m(\mathbf{sink}) = 0$$
$$m(w) = \frac{1}{n(w)} \sum_i n(v_i) \left( m(v_i) + 1 \right)$$ \\
\hline

Variance of length of paths to $\mathbf{sink}$ ($v$) &
$$v(\mathbf{sink}) = 0$$
$$v(w) = \frac{1}{n(w)} \sum_i n(v_i) \left( 
v(v_i) + \left(  m(v_i) + 1 - m(w)  \right)^2
\right)$$ \\
\hline

\end{tabular}
\end{center}
\caption{Recurrence relations for GDG-related metrics computation. $w$ is the parent and the $v_i$ are the children. The mean and variance metrics recurrence relations use formulas for combined (pooled) mean and variance of multiple sample sets.}
\label{table:rec_gdg}
\end{table}

\section{Clustering settings and examples}

\label{app:3}

\begin{figure}[h]
    \centering
    \small
    \begin{minipage}{0.45\textwidth}
        \centering
        \captionof{table}{K-Means settings: Single-core architecture}
        \begin{tabular}{|c|c|c|}
            \hline
            Cluster & n\_clusters & Silhouette coef. \\
            \hline
            size & 5 & 0.473\\
            \hline
            0 & 4 & 0.312 \\
            \hline
            1 & 6 & 0.290 \\
            \hline
            2 & 7 & 0.319 \\
            \hline
            3 & 1 & - \\
            \hline
            4 & 7 & 0.352 \\
            \hline
        \end{tabular}
    \end{minipage}
    \hfill
    \vspace{0.3cm}
    \begin{minipage}{0.45\textwidth}
        \centering
        \captionof{table}{K-Means settings: Multi-core architecture}
        \begin{tabular}{|c|c|c|}
            \hline
            Cluster & n\_clusters & Silhouette coef. \\
            \hline
            size & 5 & 0.473\\
            \hline
            0 & 6 & 0.386 \\
            \hline
            1 & 6 & 0.342 \\
            \hline
            2 & 10 & 0.349 \\
            \hline
            3 & 1 & - \\
            \hline
            4 & 6 & 0.288 \\
            \hline
        \end{tabular}
        \label{tab:clustering_settings}
    \end{minipage}
\end{figure}

\begin{figure}[h]
	\includegraphics[width=\linewidth]{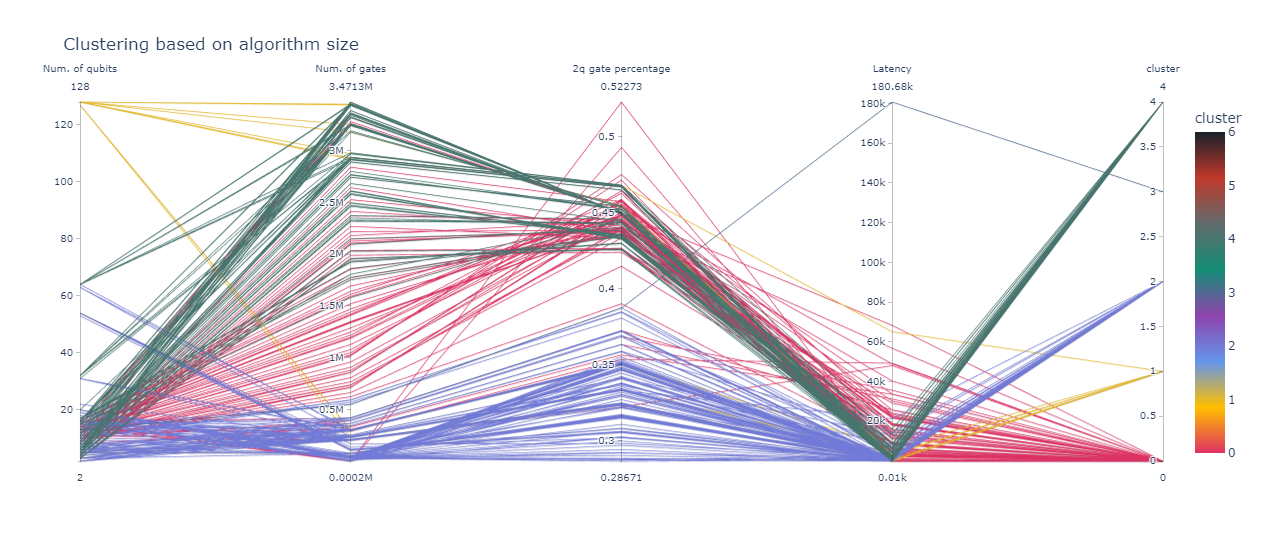}
	\caption{Clustering of quantum circuits based on size-related parameters.}
 	\label{fig:kmeans}
\end{figure}

\begin{figure}[h!]
	\centering
	\includegraphics[width= \linewidth]{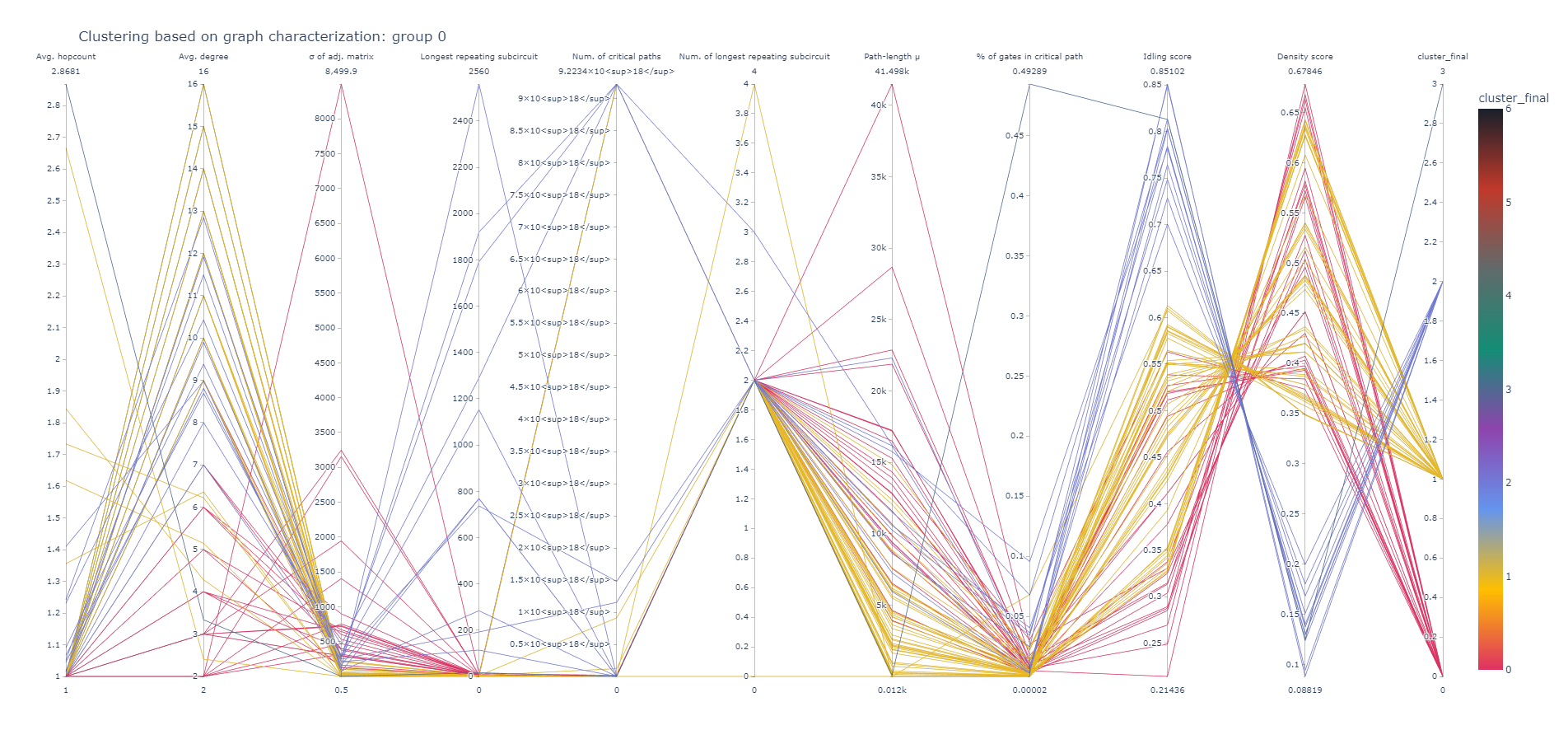}
	\caption{Sub-clustering of quantum circuits of cluster 0 (Fig. \ref{fig:kmeans}) based on other structural parameters.}
 	\label{fig:kmeans2}
\end{figure}

To cluster the benchmarks according to the specified parameters, we utilize K-Means, a centroid-based clustering algorithm \cite{1056489}. The configuration of this algorithm is detailed in Table \ref{tab:clustering_settings}, where the first column indicates the number of size-based clusters, the second column represents the number of structure-based sub-clusters within each cluster, and the third column showcases their respective Silhouette coefficients, guiding our selection process. An illustration of a cluster and its sub-clusters is provided in Figs. \ref{fig:kmeans} and \ref{fig:kmeans2} for reference.

\section{Software availability}
\label{app:4}
The online GitHub repository \noindent\href{https://github.com/QML-Group/QuantumCircuitProfiling}{https://github.com/QML-Group/QuantumCircuitProfiling} contains the following:

\begin{enumerate}
    \item Benchmarks QASM files;
    \item Code for extracting all circuit parameters and clustering together with prerequisites;
    \item Code for simulations for different compilers and architectures;
    \item Tables containing results: circuit parameters and compilation for all selected benchmarks; 
    \item Tables containing final clusters of benchmarks; and
    \item Plots of other extracted results.
\end{enumerate}

\end{document}